\documentclass[pre,superscriptaddress,tightenlines,twocolumn,showpacs]{revtex4-1}
\usepackage{graphicx,amssymb,amsmath,epsf,bm}
\usepackage[normalem]{ulem}
\usepackage{stackrel}
\usepackage{color}

\usepackage{color}

\definecolor{nblue}  {RGB}{28,130,185}

\definecolor{cgreen}  {RGB}{76,153,0}

\usepackage{latexsym,amsmath,amsfonts,tabularx,amssymb,bm,empheq}
\usepackage{graphicx,epstopdf}
\usepackage{xspace}

\usepackage{hyperref}

\newcommand{\bea}{\begin{eqnarray}}
\newcommand{\eea}{\end{eqnarray}}

\def\q{\mathbf{q}}
\def\n{\mathbf{n}}
\def\x{\textbf{x}}
\def\y{\textbf{y}}
\newcommand{\DD}{\mathcal{D}}

\def\simge{\mathrel{%
   \rlap{\raise 0.511ex \hbox{$>$}}{\lower 0.511ex \hbox{$\sim$}}}}
\def\simle{\mathrel{
   \rlap{\raise 0.511ex \hbox{$<$}}{\lower 0.511ex \hbox{$\sim$}}}}

\def\simle{\mathrel{
   \rlap{\raise 0.511ex \hbox{$<$}}{\lower 0.511ex \hbox{$\sim$}}}}

\def\simge{\mathrel{%
    \rlap{\raise 0.511ex \hbox{$>$}}{\lower 0.511ex \hbox{$\sim$}}}}

\usepackage{subfigure}
\usepackage{color}

\usepackage{tabularx}

\begin{document}

\title{
Critical probability distributions of the order parameter from the functional renormalization group
}

\author{I. Balog}
\affiliation{Institute of Physics, Bijeni\v{c}ka cesta 46, HR-10001 Zagreb, Croatia}
\author{A. Ran\c con}
\affiliation{
Univ.Lille, CNRS, UMR 8523 -- PhLAM -- Laboratoire de
Physique des Lasers Atomes et Mol\'ecules, F-59000 Lille, France
}
\author{B. Delamotte}
\affiliation{
Laboratoire de Physique Th\'eorique de la Mati\`ere Condens\'ee, UPMC,
CNRS UMR 7600, Sorbonne Universit\'e, 4, place Jussieu, 75252 Paris Cedex 05, France
}

\date{\today}

\begin{abstract}
 We show that the functional renormalization group (FRG) allows for the calculation of the probability distribution function of the sum of strongly correlated random variables.
On the example of the  three-dimensional Ising model at criticality and using the simplest implementation of the FRG, we compute the probability distribution functions  of the order parameter or equivalently its logarithm, called the rate functions in large deviations theory. 
 We compute the entire family of universal scaling functions, obtained in the limit where the system size $L$ and the correlation length of the infinite system $\xi_{\infty}$ diverge, with the ratio $\zeta=L/\xi_{\infty}$ held fixed. It compares very accurately with numerical simulations. 
\end{abstract}

\maketitle

In many different fields of research, mathematicians, physicists and even specialists of quantitative finance have paid considerable attention to the probability distribution of the sums of random variables. Here the Central Limit Theorem (CLT) plays a crucial role \cite{FellerBook, Botet_book}. It asserts that, given a large number $N$ of independent identically distributed random variables $\hat \sigma_i$ with zero mean and finite variance, their sum $\hat S=\sum_i\hat\sigma_i$ has fluctuations of order $\sqrt{N}$, and the asymptotic probability distribution function (PDF) of $\hat S/\sqrt{N}$ is a Gaussian law with finite variance. Most importantly, this result is independent of the probability law of the $\hat \sigma$'s, and the normal distribution plays the role of an attractor for the addition of an increasing number of random variables. The Gaussian distribution is therefore said to be stable and this is the most basic manifestation of what physicists call universality. The CLT has been generalized to the case where either the mean or the variance of the $\sigma_i$-law diverges: In this case, once it has been properly normalized, $\hat S$ is distributed according to one of the celebrated L\'evy-stable laws \cite{levy1954, gnedenko54,levy_book} that generalize the Gaussian law of the CLT. 

The CLT can also be generalized to situations where the $\hat \sigma_i$ are correlated \cite{Dedecker2007,Botet_book}. 
If the correlation matrix $G_{ij}$ decays sufficiently fast with a given ``distance'' $r_{ij}$ between $\hat\sigma_i$ and $\hat\sigma_j$, such that $\sum_iG_{ij}$ is finite in the limit $N\to\infty$, the correlations are said to be weak. Then, the system behaves as if it were made of uncorrelated finite size clusters of  $\hat\sigma_i$ and $\hat S$ still has fluctuations of order $\sqrt{N}$. The CLT applies again, and the distribution of $\hat S/\sqrt{N}$ is still Gaussian.

\begin{figure}[t!]
    \centering
    \includegraphics[width=8cm]{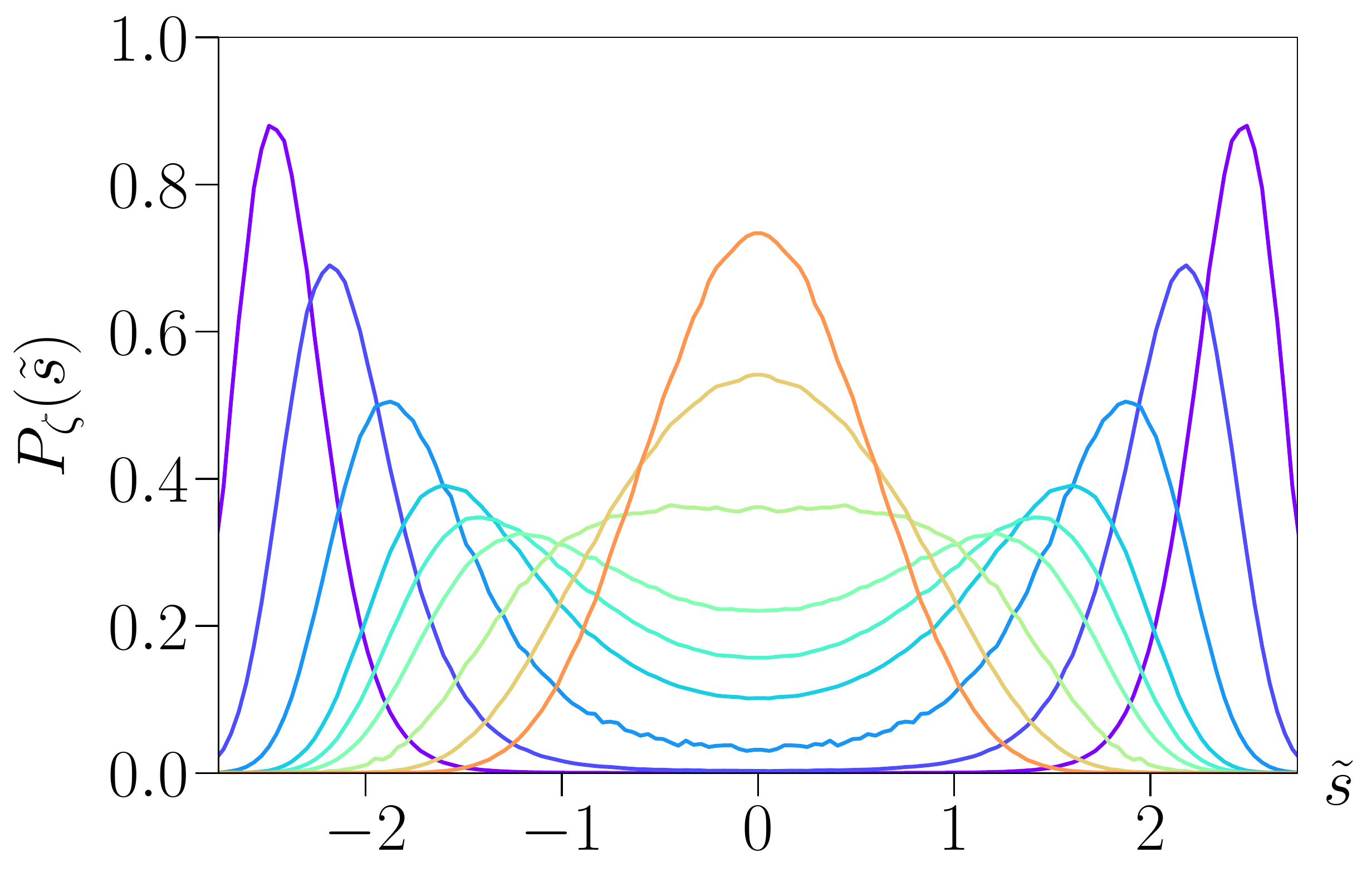}
    \caption{Different critical PDFs of the 3$d$ Ising model as functions of $\tilde s=L^{\frac{d-2+\eta}{2}}s$, obtained from Monte-Carlo simulations with periodic boundary conditions with $L=128$ for various $\zeta={\rm sgn}(T-T_c)L/\xi_{\infty}(|T-T_c|)$, with $\zeta=-4$ to $4$ by step of one (from bottom curve to top curve at the center).}
    \label{fig:my_PMC}
\end{figure} 

On the other hand, when $\sum_i G_{ij}$ diverges as $N\to\infty$, the fluctuations of $\hat S/\sqrt{N}$ also diverge and the $\hat\sigma_i$ are said to be strongly correlated. Such situations are encountered in many different contexts, from critical systems, to out-of-equilibrium dynamics such as disease propagation, surface growth or turbulence.
Our understanding of stable laws is much scarcer in this case.  Nevertheless, it is reasonable to assume that properly normalized, $\hat S/f(N)$ should here again follow a stable law. Assuming universality, these laws, that are neither Gaussian nor L\'evy, should depend only on a small number of parameters, such as the dimension of the system and its symmetries. These stable laws for strongly correlated variables have been observed experimentally or estimated numerically  with relative ease \cite{Binder1981,Binder1981a,Bruce1992,Nicolaides1988,Tsypin1994,Alexandrou1999,Tsypin2000,BHP_1,BHP_2,Portelli_02,Takeuchi2010,Martins2012,Malakis2014,Xu2020}. On the theoretical side, a few exact results have been obtained in some specific models \cite{Dyson1969,Bleher1973,ColletBook,Bleher1987,Brankov1989,Antal2004,Sasamoto2010,WuBook,Camia16}.  In generic models, the connections between CLT, stable laws and  the fixed points of the Renormalization Group (RG) have been identified \cite{JonaLasino1975,Gallavotti1975,Cassandro1978} since the early days of the Kadanoff-Wilson version of the RG \cite{WilsonKogut_74}. However, it appears that these connections have remained at the conceptual level and have not been transformed into a set of techniques for calculating PDFs applicable to strongly correlated
systems but in isolated cases with {\sl ad hoc} methods \cite{Bruce1979,Eisenriegler1987,Hilfer1993,Hilfer1995,Esser1995,Bruce1997,Rudnick1998}.
Furthermore, the connection between RG and CLTs raises two paradoxes: 1) the PDF --being an observable-- is RG-scheme independent, whereas fixed points are not; 2) as discussed below, there is a family of critical rate functions, indexed by a real number $\zeta$, but only one RG fixed point.
We show here that the functional RG (FRG) in its modern version \cite{Berges_02,Dupuis2021} is the right framework to solve these paradoxes and compute quantitatively the PDF of strongly correlated random variables.

Let us briefly review the concepts fleshed out above in the context of the Ising model in the vicinity of its second order phase transition, on which we will focus from now on.
The Hamiltonian of the ferromagnetic Ising model is $H=-J \sum_{\langle i j \rangle}\hat\sigma_i\hat\sigma_j $  with  $J>0$, $\hat\sigma_i=\pm 1$ and $\langle i j \rangle$ label nearest neighbor sites on a hypercubic $d$-dimensional lattice of linear size $L$  with periodic boundary conditions. A second order phase transition occurs in the Ising model at some finite temperature $T_c$ in $d>1$   (we focus on the non-mean-field case $d<4$). At fixed temperature $T\gtrsim T_c$ and for $r_{ij}\gg a$, $a$ the lattice spacing, the correlation function of the spins behaves as $G_{ij}\sim r_{ij}^{-d+2-\eta}\,\exp(-r_{ij}/\xi_{\infty})$ where $\eta\geq 0$ is the anomalous dimension of the spin field, and $\xi_{\infty}$ the correlation length of the infinite system (at zero magnetic field), which diverges at the transition as $|t|^{-\nu}$, $t=T-T_c$. The condition of weak correlations is thus equivalent to the finiteness of $\xi_{\infty}$.

 We are interested below in the PDF of the normalized total spin defined as $\hat s=L^{-d}\sum_i \hat\sigma_i$, the average of which is the magnetization $m=\langle \hat s\rangle$.
The fluctuations of $\hat s$ are measured by  $\langle \hat s^2\rangle$: $\langle \hat s^2\rangle= L^{-d}\,\chi$, defining the magnetic susceptibility $\chi$. For fixed $T\gtrsim T_c$,  $\chi\sim\xi_{\infty}^{2-\eta}$ independent of $L$ for $L\gg \xi_{\infty}$.
This implies that the fluctuations of $\hat S/\sqrt{N}$ are of order one: The system is weakly correlated. A precise calculation of the PDF is obtained from a saddle point approximation that becomes asymptotically exact when $L\to\infty$. As expected, it shows that the CLT  holds and that the PDF becomes indeed Gaussian in this limit: $P(\hat s=s)\propto \exp(-\frac{L^d}{2\chi}s^2)$ for $T>T_c$ and $L\to\infty$ (at fixed  $sL^{d/2}$)] \cite{ZinnForChildren}.

The argument above collapses at $T_c$  and fixed $L$, where $\xi_{\infty}~\gg~L$, because $\chi$ scales with $L$ as: $\chi\sim\int^L d^dr\, r^{-d+2-\eta}\sim L^{2-\eta}$ which diverges when $L~\to~\infty$. This implies that $\langle \hat s^2\rangle\sim L^{-d+2-\eta}$ and that the fluctuations of $\hat S/\sqrt{N}$  diverge as $L^{\frac{2-\eta}{2}}=N^{\frac{2-\eta}{2d}}$.
The spins are strongly correlated and the standard CLT does no longer hold: 
 The saddle-point approximation fails and $P$ has no reason to be a Gaussian anymore. 
However, the scaling of the  fluctuations of $\hat s$ suggest that $P$ is a universal function of the scaling variable $\tilde s=s\, L^{\frac{d-2+\eta}{2}}$ \cite{Bouchaud1990}.

 It is rarely stressed that there is not only one PDF at criticality, but an infinity corresponding to the inequivalent ways to take the limit $L\to\infty$ and $T\to T_c^+$, i.e., $\xi_{\infty}\to\infty$, see Fig. \ref{fig:my_PMC} \footnote{The shapes of the family of PDFs also depend crucially on the boundary conditions \cite{Binder1981a}. Here we focus on periodic boundary conditions only.}.
 Indeed, choose any sequence $T_L>T_c$ converging to $T_c$, such that $\zeta=L/\xi_{\infty}(T_L)$ is constant. Then,  for instance if  $\xi_{\infty}(T_L)\ll L$ and from the discussion above, $\langle \hat s^2\rangle\sim L^{-d}\,\chi\sim L^{-d+2-\eta}\zeta^{\eta-2}$.
 Once again, and even though $\xi_{\infty}(T_L)$ is finite at any $L$, the spins become more and more strongly correlated as $T_L\to T_c$. Therefore the PDF must be  non-trivial for all values of $\zeta$ even in the limit $\zeta\gg1$ (i.e. $\xi_{\infty}(T_L)\ll L$). In this limit, we expect to recover some Gaussian-like features for typical values of $s$ because the system looks for all $T_L>T_c$ as a  collection of uncorrelated small blocks of spins of sizes $\xi_{\infty}(T_L)$. However, some non-gaussianity should remain in the tails of the PDF reminiscent of the strong correlations present at criticality where $\chi$ is diverging. 

Assuming scaling, the PDF must depend on $\xi_{\infty}$ and $L$ only through the ratio $\zeta$, which we parametrize as
\begin{equation}
    P_\zeta(\hat s=s) \approx  e^{-L^d I(s,\xi_{\infty},L)}  \approx  e^{-I_\zeta(\tilde{s})}.
    \label{PDF-zeta}
\end{equation}
This relation defines the rate function $I(s,\xi_{\infty},L)$, as it is known in large deviations theory   (also known as the ``constraint effective potential'' in quantum field theory  \cite{Fukuda1975,ORaifeartaigh1986,Gockeler1991}), as well as its scaling function $I_\zeta(\tilde s)$. Notice that we could define as well a family of universal critical PDFs when coming from the low temperature phase, $T\to T_c^-$. To tackle both cases at once, we define $\zeta={\rm sgn}(t)L/\xi_{\infty}(|t|)$. We show in Fig.~\ref{fig:my_PMC} some of these PDFs obtained numerically (see below) in $d=3$ with periodic boundary conditions.

These probabilistic arguments do not allow for computing $I_\zeta(\tilde{s})$.  In the following we show that FRG yields a general formalism for such calculations. Being interested in universal PDFs, we replace the lattice Ising model by a $\mathbb Z_2$-invariant field theory for which $\hat s=L^{-d}\int_\x \hat\phi(\x)$ and thus:
\begin{equation}
P(\hat s= s) =\mathcal N\int \DD\hat\phi\, \delta\left(s-\hat s\right)\exp(-{\cal H}[\hat\phi]),
\label{PDF-L}
\end{equation}
with $\mathcal N$ a normalization factor. Noting that the delta-function can be  replaced by a infinitely peaked Gaussian, $\delta(z)\propto \exp(-M^2 z^2/2)$ with $M^2\to\infty$, the PDF can be interpreted as the partition function ${\cal Z}_M$ of a system with Hamiltonian ${\cal H}_M[\hat\phi]={\cal H}[\hat\phi]+\frac{M^2}{2} \left(\int_\x (\hat\phi(\x)-s)\right)^2$, that is, $P(s)\propto \lim_{M\to\infty} \mathcal Z_M$. Remark that $M=0$ corresponds to the standard partition function ${\cal Z}$ of the model (at finite size $L$). 

For a critical theory, the computation of ${\cal Z}_M$ is plagued with the singularities induced by the long-distance/small-wavenumber fluctuations. The modern version of the FRG, tailored to deal with this difficulty \cite{Wetterich1991, Wetterich1993, Wetterich1993a}, consists in freezing out these modes in the partition function while leaving unchanged the others and by gradually decreasing to zero the scale $k$ that separates the low and high wavenumber modes: This generates the RG flow of partition functions or equivalently, of Hamiltonians.  

A one-parameter family of models with partition functions ${\cal Z}_{M,k}[h]$ is thus built by changing the original Hamiltonian ${\cal H}_M$ into ${\cal H}_M+\Delta {\cal H}_k- h.\hat\phi$, where $h$ is a magnetic field (or source) and the dot in $h.\hat\phi$ implies an integral over space or momentum: ${\cal Z}_{M,k}[h]=\int \DD\hat\phi \,\exp(-{\cal H}_M-\Delta {\cal H}_k + h.\hat\phi)$.  Here, $\Delta {\cal H}_k$ is the term designed to effectively freeze the low wavenumber fluctuations $\hat\phi(\vert\q\vert<k)$ while leaving unchanged the high wavenumber modes $\hat\phi(\vert\q\vert>k)$. It is chosen quadratic: $\Delta {\cal H}_k=1/2\, \hat \phi.R_k.\hat \phi$ with $R_k(\x,\y)$ such that (i) when $k\sim a^{-1}$, $R_{k\sim a^{-1}}(\vert\q\vert)$ is very large for all $\vert\q\vert$ which implies that all fluctuations are frozen; and (ii) $R_{k=0}(\vert\q\vert)\equiv 0$ so that all fluctuations are integrated over and ${\cal Z}_{M,k=0}[h]={\cal Z}_M[h]$. Varying the scale $k$ between $a^{-1}$ and 0 induces the RG flow of ${\cal Z}_{M,k}[h]$, in which fluctuations of wavenumbers $\vert \q\vert>k$ are progressively integrated over. 

Actual calculations of ${\cal Z}_{M,k}[h]$ require to perform approximations that are known to be controlled only when working with the (slightly modified) Legendre transform of $\log{\cal Z}_{M,k}[h]$ with respect to $h$, $\Gamma_{M,k}[\phi]$ \cite{PhysRevB.68.064421,Balog2019,PhysRevE.101.042113}, defined as
\begin{equation}
\Gamma_{M,k}[\phi]=-\ln \mathcal Z_{M,k}[h]+h.\phi-\frac12\phi. R_k.\phi-\frac{M^2}{2} \{\phi-s\}^2,
\label{eq_def_GammaA_LT_text}
\end{equation}
where $\{\phi-s\}^2\equiv\left(\int_\x (\phi(\x)-s)\right)^2$ and $\phi(\x)=\langle\hat\phi(\x)\rangle=\frac{\delta { \mathcal Z}_{M,k}}{\delta h(\x)}$.
It can also be defined as (see Suppl. Mat. \cite{SM})
\begin{equation}
e^{-\Gamma_{M,k}[\phi]}=\int \DD\hat\phi\, e^{-{\cal H}[\hat\phi] -\frac12 (\hat\phi-\phi).R_{M,k}.(\hat\phi-\phi) +\frac{\delta\Gamma_{M,k}}{\delta\phi}.(\hat\phi-\phi)},
\label{second-def-gammak}
\end{equation}
where $R_{M,k}(\x,\y)=R_k(\x,\y)+M^2$, or in momentum space $R_{M,k}(\q)=R_k(\q)+M^2\delta_{\q,0}$, with $\q=\frac{2\pi}{L}\n$ and $\n\in \mathbb Z^d$. Eq. \eqref{second-def-gammak} has the advantage of explicitly showing that $\Gamma_{M,k}$ does not depend on $s$. Up to the replacement of $R_k$ by $R_{M,k}$, $\Gamma_{M,k}$ is formally identical to the usual scale-dependent  effective action $\Gamma_k$ introduced in FRG \cite{Dupuis2021}, and indeed, $\Gamma_{k}[\phi]=\Gamma_{M=0,k}[\phi]$. The exact RG equation satisfied by $\Gamma_{M,k}[\phi]$ is the usual Wetterich equation in the presence of the regulator $R_{M,k}$:
\begin{equation}
\partial_k \Gamma_{M,k}[\phi]=\frac{1}{2}\int_{\x,\y}\partial_k R_{M,k}(\x,\y)\left(\Gamma_{M,k}^{(2)}+R_{M,k}\right)^{-1}(\x,\y),
\label{eq_Wetterich}
\end{equation} 
where $\Gamma_{M,k}^{(2)}=\Gamma_{M,k}^{(2)}[\x,\y;\phi]=\frac{\delta^2\Gamma_{M,k} }{\delta\phi(\x)\delta\phi(\y)}$.

\begin{figure}[t!]
    \centering
    \includegraphics[width=8cm]{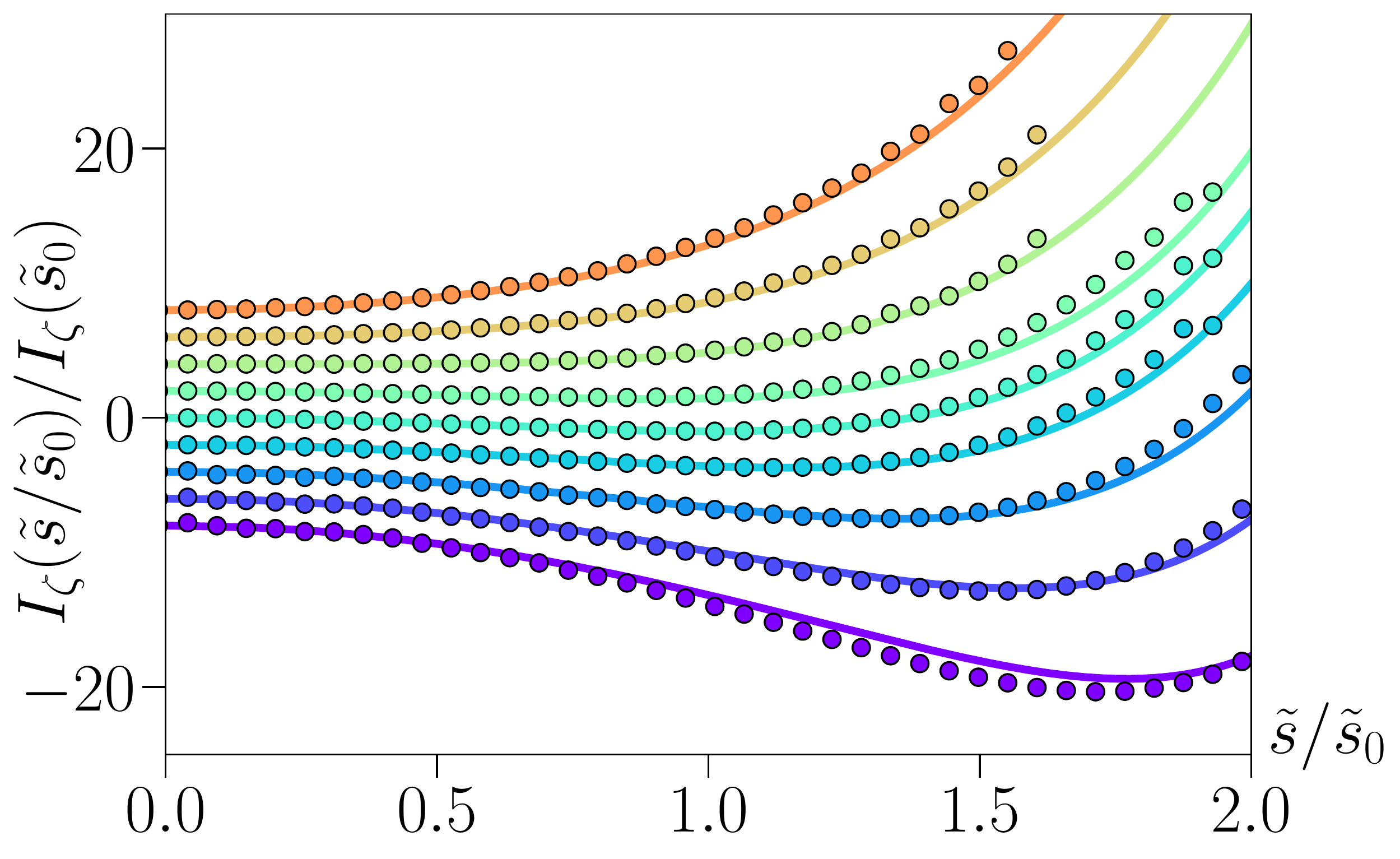}
    \caption{Normalized rate functions $I_\zeta(\tilde s)$ of the 3$d$ Ising model obtained from FRG (full line) and MC simulations (symbols) performed on the cubic lattice with periodic boundary conditions for $\zeta=-4$ to $4$ (from bottom to top, same color code as in Fig.~\ref{fig:my_PMC}). The normalization point $\tilde s_0$ is the position of the minimum of $I_{\zeta=0}$. The rate functions have been shifted for better visibility.
    }
    \label{fig_comp}
\end{figure} 

Defining $\check\Gamma_k[\phi]=\lim_{M\to\infty}\Gamma_{M,k}[\phi]$, the additional $k$-independent term $M^2\delta_{\q,0}$ completely freezes the zero-momentum mode $\int_\x \hat\phi(\x)$ in $\check\Gamma_k[\phi]$, and we  show in \cite{SM} that when evaluated in constant field $\phi(\x)=s$, $L^{-d}\check\Gamma_{k}[s]=I_k(s)$ is a scale dependent rate function such that  $P(s)\propto \lim_{k\to 0} \exp(-L^dI_k(s))$. [In contrast, when evaluated in a constant field $\phi(\x)=m$, the effective action $\Gamma_{k}[\phi=m]$ is $L^d U_k(m)$ where $U_k(m)$ is the $k$-dependent effective potential that becomes the true effective potential at $k=0$. 
In particular, $\Gamma[\phi]=\Gamma_{k=0}[\phi]$ being the Legendre transform of $\ln {\cal Z}[h]$, the effective potential $U(m)=U_{k=0}(m)$ is a convex function of $m$ \cite{Dupuis2021}. Note that both $\Gamma$ and $I$ are RG-scheme independent by construction.]
Our aim in the following is to compute $\check\Gamma_k[\phi=s]$ and to evaluate it at $k=0$. For this purpose, we now study the flow of  $\check\Gamma_k$ comparing it with the better known flow of $\Gamma_k$.

For $\zeta\ll 1$ and $a^{-1}\gg k\gg 1/L\gg 1/\xi_{\infty}$,  the regulator $R_{k}$ effectively freezes the zero-momentum mode in $\Gamma_k$, which makes its flow identical to that of $\check\Gamma_k$, up to corrections of order $(kL)^{-d}$. In this range of $k$, the system is self-similar because both $a$ and $L$ play no role in the flows. It follows that both $U_k$ and $I_k$ obey a scaling form $I_k(\phi)\simeq U_k(\phi)=k^{d}\,\tilde U^*(\phi\, k^{-(d-2+\eta)/2})$, where $\tilde U^*$ is $k$-independent, that is, it is the dimensionless fixed point potential of the RG flow of $\Gamma_k$ \cite{Dupuis2021}. It is a non-convex function that has the typical double well form, see below.

When $k$ becomes of order $2\pi/L$, the flows of $U_k$ and $I_k$ start to differ significantly. On the one hand, the flow of $U_k$ becomes essentially that of a zero-dimensional system (corresponding to the fluctuations of the zero-momentum mode only), and $\lim_{k\to 0}U_k(m)$  becomes convex with a curvature at small $m$ given by $\chi^{-1}\propto L^{-2+\eta}$. On the other hand, the flow of $I_k(s)$ stops typically for $k\lesssim 2\pi/L$ because in this quantity the zero-momentum mode is frozen by the $M\to \infty$ term. In particular, this allows for a non-convex shape of $I(s)=I_{k= 0}(s)$ and $L^d I(s)$ is found to naturally be a function of $\tilde s =s L^{\frac{d-2+\eta}{2}}$.

The above picture is modified when $\zeta\gg 1$ ($T>T_c$, $L\gg\xi_\infty$), because the system size can no longer play any significant role  when $\xi_{\infty}\ll L$. In particular, the flows  of $U_k$ and $I_k$ rapidly stop for $k\lesssim 1/\xi_{\infty}$ and it makes no difference whether the zero mode is completely frozen or not. Approaching criticality from  the disordered phase, we therefore find that $I_{k=0}(s)\simeq U_{k=0}(m=s)$. These functions are  convex with positive curvature $\chi^{-1}\propto \xi_{\infty}^{-2+\eta}$ at $s=0$. Working at fixed $\zeta$, we thus have $L^d I_{k=0}(s)\propto \zeta^{2-\eta}L^{d-2+\eta}s^2$ at small $s$. The PDF is therefore Gaussian at small $s$ as in the CLT, which is expected because the system looks like a collection of uncorrelated clusters of spins of extension $\xi_{\infty}$. However, since the susceptibility diverges at $T=T_c$ as $\xi_{\infty}^{-2+\eta}$, the fluctuations are anomalously large  compared with the usual CLT because they are of order $L^{-\frac{d-2+\eta}{2}}$ instead of the standard $L^{-d/2}$.
Varying $\zeta$  then generates a smooth family of rate functions, the shapes of which depend on the competition between $L$ and $\xi_{\infty}$ in the flow. Furthermore, $I_\zeta(\tilde s)$ behaves as $\tilde s ^{\frac{2d}{d-2+\eta}}$ at large $\tilde s$, a behavior inherited from $\tilde U^*$ \cite{SM}.

\begin{figure}[!t]
    \centering
    \includegraphics[width=8cm]{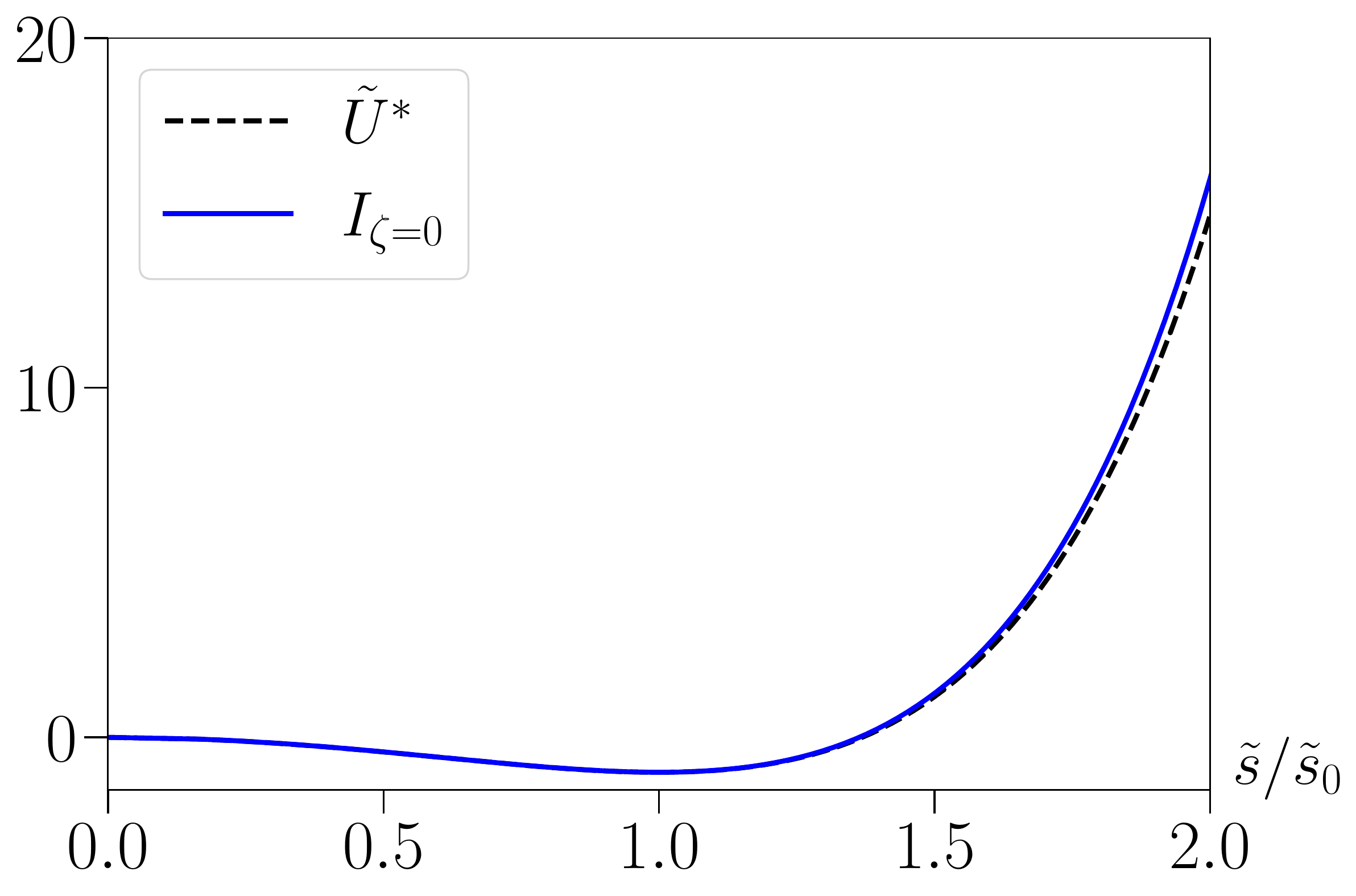}
    \caption{Scaling rate function at $\zeta=0$ obtained from FRG (blue) as a function of $\tilde s$, and fixed point potential obtained with the same regulator as function of $\tilde \phi=\tilde s$. Both have been normalized such that their minimum is $-1$ at $1$. The difference is only visible in the tail. }
    \label{fig_comp_PP_FP}
\end{figure}

To compute in practice the rate function and find its specific shape depending on $\zeta$, it is necessary to solve the flow equation Eq.~\eqref{eq_Wetterich}. This cannot be done exactly, and it is necessary to perform approximations. Here, we focus on the simplest of such approximations which nevertheless allows for a functional calculation of the rate function, the so-called Local Potential Approximation (LPA). It amounts to using the Ansatz
$
    \check\Gamma_k[\phi]=\int_\x\left(\frac{1}{2}(\partial \phi)^2+I_k(\phi(\x))\right)
$,
and projecting the flow equation onto this Ansatz. The corresponding LPA flow equation is then closed for the scale-dependent rate function and reads
\begin{equation}
\partial_k I_{k}(s)=\frac{1}{2L^d}\sum_{\q\neq 0}\frac{\partial_k R_{k}(\q)}{\q^2+R_k(\q)+I''_{k}(s)},
\label{flow_I}
\end{equation} 
and we use the ``exponential regulator'' $R_k(\q)=\alpha k^2 e^{-\q^2/k^2}$ with $\alpha=4.65$ \cite{SM}.
Note that at LPA the anomalous dimension vanishes, $\eta=0$, but since $\eta\ll 1$ for the three-dimensional Ising model, we expect the approximation to correctly capture the shape of the rate function.
The scaling functions $I_\zeta(\tilde s)=I_{\zeta,k=0}(\tilde s)$ obtained from integrating the LPA flow, Eq.~\eqref{flow_I}, are shown as solid lines in Fig.~\ref{fig_comp} for various $\zeta$, see \cite{SM}. 
We have verified that the resulting rate functions obey the expected scaling, are functions of $\tilde s$ and  $\zeta$ only, and only very weakly depend on the regulator function $R_k$ \cite{SM}.

In the same figure, we compare our FRG results to the rate functions obtained from Monte-Carlo (MC) simulations on the cubic lattice with periodic boundary conditions, using a Wolff algorithm \cite{Wolff1989} with histogram reweighting \cite{Ferrenberg1988}, also used to generate Fig.~\ref{fig:my_PMC}, see  \cite{SM}.Since lattice and field theory calculations use different units, it requires rescalings of the $x$-axis (magnitude of the total spin $s$) and $y$-axis (associated to the microscopic length scales, since $I(s)$ is a density) in the plot of $I(s)$, Fig.~\ref{fig_comp}.
 Importantly, these model-dependent lengths are independent of $\zeta$, and should be determined from only one value of $\zeta$ (we use $\zeta=0$). We find that to compare the rate function obtained from MC for a given $\zeta_{\rm MC}$ to that obtained from LPA at $\zeta_{\rm LPA}$ necessitates a rescaling of $\zeta_{\rm LPA}$, with $\zeta_{\rm MC}\simeq 0.9\, \zeta_{\rm LPA}$ \cite{SM}.  We attribute this to errors in the computation of $\xi_{\infty}$ induced by LPA. With this small caveat, we find a very good agreement between simulations and FRG on the whole range of $\zeta \in [-4,4]$.
 Note that the rate functions become strictly convex for $\zeta\gtrsim 2.2$.

 It is interesting to note that $I_{\zeta=0}$ is very similar to the fixed point potential, when properly normalized, see. Fig.~\ref{fig_comp_PP_FP}. This could explain why the fixed point of the RG has long been thought to describe the critical PDFs \cite{Bruce1979,Esser1995,Rudnick1998}. However, this cannot be true exactly because the dependence of the fixed point effective potential $\tilde U^*$ on the choice of regulator $R_k$ cannot be normalized out. This can be shown explicitly in the large $N$ limit of the critical $O(N)$ model \cite{next_paper_2035}. Reciprocally, our work confirms that RG is deeply related to probability theory since computing a fixed point is actually almost synonymous to computing the $I_{\zeta=0}$  but for the zero-mode which is excluded in the rate function.   This elucidates the long-standing paradoxes arising from the confusion between the fixed point potential $\tilde U^*$ and $I_{\zeta=0}$ which, although very closely related, are conceptually different.

Our work raises many questions and paves the way to many applications that we want to briefly review below. For instance, the method can be generalized to all pure statistical systems at thermal equilibrium, with probably very good results at least when the LPA is accurate, that is, when $\eta$ is small. 
The generalization to disordered and/or out of equilibrium systems, where very little is known about the computation of critical PDFs, certainly requires to adapt the formalism. This should be feasible since FRG already yields fairly accurate results for such problems like the random field O(N) models \cite{PhysRevLett.93.267008,PhysRevLett.107.041601,PhysRevB.85.104202}, reaction-diffusion models \cite{PhysRevLett.92.195703,PhysRevLett.95.100601} and the Kardar-Parisi-Zhang equation \cite{PhysRevLett.104.150601,PhysRevE.84.061128,PhysRevE.89.022108,PhysRevE.93.063101} to mention just a few \cite{Dupuis2021}. Also, the coexistence region in the low-temperature phase is highly non-trivial, scaling as a surface term, and necessitates to go beyond LPA, which does not capture domain walls. This could explain why our results do not agree as well with MC simulations for large and negative $\zeta$.
However, the LPA can be systematically improved via a derivative expansion or the Blaizot-Mendez-Wschebor approximation scheme \cite{Blaizot2006,Benitez2009}.
The study of the convergence along the lines of \cite{Balog2019,PhysRevE.101.042113} for the rate function is left for future work.

\acknowledgements
We thank  O. B\'enichou, N. Dupuis, G. Tarjus and  N. Wschebor for discussions and feedbacks. BD thanks F. Benitez, M. Tissier and Z. R\'acz for many  discussions in an early stage of this work. AR also thanks G. Verley for discussions on this and related subjects. IB acknowledges the support of the the QuantiXLie Centre of Excellence, a project cofinanced by the Croatian Government and European Union through the European Regional Development Fund - the Competitiveness and Cohesion Operational Programme (Grant KK.01.1.1.01.0004). BD acknowledges the support from the French ANR through the project NeqFluids (grant ANR-18-CE92-0019).
AR is supported by the Research Grants QRITiC I-SITE ULNE/ ANR-16-IDEX-0004 ULNE.

\bibliography{CLT_paper}

\begin{thebibliography}{69}%
\makeatletter
\providecommand \@ifxundefined [1]{%
 \@ifx{#1\undefined}
}%
\providecommand \@ifnum [1]{%
 \ifnum #1\expandafter \@firstoftwo
 \else \expandafter \@secondoftwo
 \fi
}%
\providecommand \@ifx [1]{%
 \ifx #1\expandafter \@firstoftwo
 \else \expandafter \@secondoftwo
 \fi
}%
\providecommand \natexlab [1]{#1}%
\providecommand \enquote  [1]{``#1''}%
\providecommand \bibnamefont  [1]{#1}%
\providecommand \bibfnamefont [1]{#1}%
\providecommand \citenamefont [1]{#1}%
\providecommand \href@noop [0]{\@secondoftwo}%
\providecommand \href [0]{\begingroup \@sanitize@url \@href}%
\providecommand \@href[1]{\@@startlink{#1}\@@href}%
\providecommand \@@href[1]{\endgroup#1\@@endlink}%
\providecommand \@sanitize@url [0]{\catcode `\\12\catcode `\$12\catcode
  `\&12\catcode `\#12\catcode `\^12\catcode `\_12\catcode `\%12\relax}%
\providecommand \@@startlink[1]{}%
\providecommand \@@endlink[0]{}%
\providecommand \url  [0]{\begingroup\@sanitize@url \@url }%
\providecommand \@url [1]{\endgroup\@href {#1}{\urlprefix }}%
\providecommand \urlprefix  [0]{URL }%
\providecommand \Eprint [0]{\href }%
\providecommand \doibase [0]{http://dx.doi.org/}%
\providecommand \selectlanguage [0]{\@gobble}%
\providecommand \bibinfo  [0]{\@secondoftwo}%
\providecommand \bibfield  [0]{\@secondoftwo}%
\providecommand \translation [1]{[#1]}%
\providecommand \BibitemOpen [0]{}%
\providecommand \bibitemStop [0]{}%
\providecommand \bibitemNoStop [0]{.\EOS\space}%
\providecommand \EOS [0]{\spacefactor3000\relax}%
\providecommand \BibitemShut  [1]{\csname bibitem#1\endcsname}%
\let\auto@bib@innerbib\@empty
\bibitem [{\citenamefont {Feller}(1971)}]{FellerBook}%
  \BibitemOpen
  \bibfield  {author} {\bibinfo {author} {\bibfnamefont {W.}~\bibnamefont
  {Feller}},\ }\href@noop {} {\emph {\bibinfo {title} {An introduction to
  Probability Theory, Vols 1 and 2}}}\ (\bibinfo  {publisher} {Wiley, New
  York},\ \bibinfo {year} {1971})\BibitemShut {NoStop}%
\bibitem [{\citenamefont {{Botet}}\ and\ \citenamefont
  {{P{\l}oszajczak}}(2002)}]{Botet_book}%
  \BibitemOpen
  \bibfield  {author} {\bibinfo {author} {\bibfnamefont {R.}~\bibnamefont
  {{Botet}}}\ and\ \bibinfo {author} {\bibfnamefont {M.}~\bibnamefont
  {{P{\l}oszajczak}}},\ }\href {\doibase 10.1142/4916} {\emph {\bibinfo {title}
  {{Universal Fluctuations: The Phenomenology of Hadronic Matter}}}}\ (\bibinfo
  {year} {2002})\BibitemShut {NoStop}%
\bibitem [{\citenamefont {L{\'e}vy}(1954)}]{levy1954}%
  \BibitemOpen
  \bibfield  {author} {\bibinfo {author} {\bibfnamefont {P.}~\bibnamefont
  {L{\'e}vy}},\ }\href {https://books.google.hr/books?id=tg3vAAAAMAAJ} {\emph
  {\bibinfo {title} {Th{\'e}orie de l'addition des variables
  al{\'e}atoires}}},\ Collection des monographies des probabilit{\'e}s\
  (\bibinfo  {publisher} {Gauthier-Villars},\ \bibinfo {year}
  {1954})\BibitemShut {NoStop}%
\bibitem [{\citenamefont {Gnedenko}\ and\ \citenamefont
  {Kolmogorov}(1954)}]{gnedenko54}%
  \BibitemOpen
  \bibfield  {author} {\bibinfo {author} {\bibfnamefont {B.~V.}\ \bibnamefont
  {Gnedenko}}\ and\ \bibinfo {author} {\bibfnamefont {A.~N.}\ \bibnamefont
  {Kolmogorov}},\ }\href@noop {} {\emph {\bibinfo {title} {Limit distributions
  for sums of independent random variables}}}\ (\bibinfo  {publisher}
  {Cambridge, Mass: Addison-Wesley Pub. Co.},\ \bibinfo {year}
  {1954})\BibitemShut {NoStop}%
\bibitem [{\citenamefont {Ibe}(2013)}]{levy_book}%
  \BibitemOpen
  \bibfield  {author} {\bibinfo {author} {\bibfnamefont {O.~C.}\ \bibnamefont
  {Ibe}},\ }in\ \href {\doibase
  https://doi.org/10.1016/B978-0-12-407795-9.00011-6} {\emph {\bibinfo
  {booktitle} {Markov Processes for Stochastic Modeling (Second Edition)}}},\
  \bibinfo {editor} {edited by\ \bibinfo {editor} {\bibfnamefont {O.~C.}\
  \bibnamefont {Ibe}}}\ (\bibinfo  {publisher} {Elsevier},\ \bibinfo {address}
  {Oxford},\ \bibinfo {year} {2013})\ \bibinfo {edition} {second edition}\
  ed.,\ pp.\ \bibinfo {pages} {329--347}\BibitemShut {NoStop}%
\bibitem [{\citenamefont {Dedecker}\ \emph {et~al.}(2007)\citenamefont
  {Dedecker}, \citenamefont {Doukhan}, \citenamefont {Lang}, \citenamefont
  {Rafael}, \citenamefont {Louhichi},\ and\ \citenamefont
  {Prieur}}]{Dedecker2007}%
  \BibitemOpen
  \bibfield  {author} {\bibinfo {author} {\bibfnamefont {J.}~\bibnamefont
  {Dedecker}}, \bibinfo {author} {\bibfnamefont {P.}~\bibnamefont {Doukhan}},
  \bibinfo {author} {\bibfnamefont {G.}~\bibnamefont {Lang}}, \bibinfo {author}
  {\bibfnamefont {L.~R.~J.}\ \bibnamefont {Rafael}}, \bibinfo {author}
  {\bibfnamefont {S.}~\bibnamefont {Louhichi}}, \ and\ \bibinfo {author}
  {\bibfnamefont {C.}~\bibnamefont {Prieur}},\ }\href {\doibase
  10.1007/978-0-387-69952-3} {\emph {\bibinfo {title} {Weak Dependence: With
  Examples and Applications}}}\ (\bibinfo  {publisher} {Springer New York},\
  \bibinfo {year} {2007})\BibitemShut {NoStop}%
\bibitem [{\citenamefont {Binder}(1981{\natexlab{a}})}]{Binder1981}%
  \BibitemOpen
  \bibfield  {author} {\bibinfo {author} {\bibfnamefont {K.}~\bibnamefont
  {Binder}},\ }\href {\doibase 10.1103/PhysRevLett.47.693} {\bibfield
  {journal} {\bibinfo  {journal} {Phys. Rev. Lett.}\ }\textbf {\bibinfo
  {volume} {47}},\ \bibinfo {pages} {693} (\bibinfo {year}
  {1981}{\natexlab{a}})}\BibitemShut {NoStop}%
\bibitem [{\citenamefont {Binder}(1981{\natexlab{b}})}]{Binder1981a}%
  \BibitemOpen
  \bibfield  {author} {\bibinfo {author} {\bibfnamefont {K.}~\bibnamefont
  {Binder}},\ }\href {\doibase 10.1007/BF01293604} {\bibfield  {journal}
  {\bibinfo  {journal} {Zeitschrift f{\"u}r Physik B Condensed Matter}\
  }\textbf {\bibinfo {volume} {43}},\ \bibinfo {pages} {119} (\bibinfo {year}
  {1981}{\natexlab{b}})}\BibitemShut {NoStop}%
\bibitem [{\citenamefont {Bruce}\ and\ \citenamefont
  {Wilding}(1992)}]{Bruce1992}%
  \BibitemOpen
  \bibfield  {author} {\bibinfo {author} {\bibfnamefont {A.~D.}\ \bibnamefont
  {Bruce}}\ and\ \bibinfo {author} {\bibfnamefont {N.~B.}\ \bibnamefont
  {Wilding}},\ }\href {\doibase 10.1103/PhysRevLett.68.193} {\bibfield
  {journal} {\bibinfo  {journal} {Phys. Rev. Lett.}\ }\textbf {\bibinfo
  {volume} {68}},\ \bibinfo {pages} {193} (\bibinfo {year} {1992})}\BibitemShut
  {NoStop}%
\bibitem [{\citenamefont {Nicolaides}\ and\ \citenamefont
  {Bruce}(1988)}]{Nicolaides1988}%
  \BibitemOpen
  \bibfield  {author} {\bibinfo {author} {\bibfnamefont {D.}~\bibnamefont
  {Nicolaides}}\ and\ \bibinfo {author} {\bibfnamefont {A.~D.}\ \bibnamefont
  {Bruce}},\ }\href {\doibase 10.1088/0305-4470/21/1/028} {\bibfield  {journal}
  {\bibinfo  {journal} {Journal of Physics A: Mathematical and General}\
  }\textbf {\bibinfo {volume} {21}},\ \bibinfo {pages} {233} (\bibinfo {year}
  {1988})}\BibitemShut {NoStop}%
\bibitem [{\citenamefont {Tsypin}(1994)}]{Tsypin1994}%
  \BibitemOpen
  \bibfield  {author} {\bibinfo {author} {\bibfnamefont {M.~M.}\ \bibnamefont
  {Tsypin}},\ }\href {\doibase 10.1103/PhysRevLett.73.2015} {\bibfield
  {journal} {\bibinfo  {journal} {Phys. Rev. Lett.}\ }\textbf {\bibinfo
  {volume} {73}},\ \bibinfo {pages} {2015} (\bibinfo {year}
  {1994})}\BibitemShut {NoStop}%
\bibitem [{\citenamefont {Alexandrou}\ \emph {et~al.}(1999)\citenamefont
  {Alexandrou}, \citenamefont {Bori\ifmmode~\mbox{\c{c}}\else \c{c}\fi{}i},
  \citenamefont {Feo}, \citenamefont {de~Forcrand}, \citenamefont {Galli},
  \citenamefont {Jergerlehner},\ and\ \citenamefont
  {Takaishi}}]{Alexandrou1999}%
  \BibitemOpen
  \bibfield  {author} {\bibinfo {author} {\bibfnamefont {C.}~\bibnamefont
  {Alexandrou}}, \bibinfo {author} {\bibfnamefont {A.}~\bibnamefont
  {Bori\ifmmode~\mbox{\c{c}}\else \c{c}\fi{}i}}, \bibinfo {author}
  {\bibfnamefont {A.}~\bibnamefont {Feo}}, \bibinfo {author} {\bibfnamefont
  {P.}~\bibnamefont {de~Forcrand}}, \bibinfo {author} {\bibfnamefont
  {A.}~\bibnamefont {Galli}}, \bibinfo {author} {\bibfnamefont
  {F.}~\bibnamefont {Jergerlehner}}, \ and\ \bibinfo {author} {\bibfnamefont
  {T.}~\bibnamefont {Takaishi}},\ }\href {\doibase 10.1103/PhysRevD.60.034504}
  {\bibfield  {journal} {\bibinfo  {journal} {Phys. Rev. D}\ }\textbf {\bibinfo
  {volume} {60}},\ \bibinfo {pages} {034504} (\bibinfo {year}
  {1999})}\BibitemShut {NoStop}%
\bibitem [{\citenamefont {Tsypin}\ and\ \citenamefont
  {Bl\"ote}(2000)}]{Tsypin2000}%
  \BibitemOpen
  \bibfield  {author} {\bibinfo {author} {\bibfnamefont {M.~M.}\ \bibnamefont
  {Tsypin}}\ and\ \bibinfo {author} {\bibfnamefont {H.~W.~J.}\ \bibnamefont
  {Bl\"ote}},\ }\href {\doibase 10.1103/PhysRevE.62.73} {\bibfield  {journal}
  {\bibinfo  {journal} {Phys. Rev. E}\ }\textbf {\bibinfo {volume} {62}},\
  \bibinfo {pages} {73} (\bibinfo {year} {2000})}\BibitemShut {NoStop}%
\bibitem [{\citenamefont {{Bramwell}}\ \emph {et~al.}(1998)\citenamefont
  {{Bramwell}}, \citenamefont {{Holdsworth}},\ and\ \citenamefont
  {{Pinton}}}]{BHP_1}%
  \BibitemOpen
  \bibfield  {author} {\bibinfo {author} {\bibfnamefont {S.~T.}\ \bibnamefont
  {{Bramwell}}}, \bibinfo {author} {\bibfnamefont {P.~C.~W.}\ \bibnamefont
  {{Holdsworth}}}, \ and\ \bibinfo {author} {\bibfnamefont {J.~F.}\
  \bibnamefont {{Pinton}}},\ }\href {\doibase 10.1038/25083} {\bibfield
  {journal} {\bibinfo  {journal} {\nat}\ }\textbf {\bibinfo {volume} {396}},\
  \bibinfo {pages} {552} (\bibinfo {year} {1998})}\BibitemShut {NoStop}%
\bibitem [{\citenamefont {{Bramwell}}\ \emph {et~al.}(2000)\citenamefont
  {{Bramwell}}, \citenamefont {{Christensen}}, \citenamefont {{Fortin}},
  \citenamefont {{Holdsworth}}, \citenamefont {{Jensen}}, \citenamefont
  {{Lise}}, \citenamefont {{L{\'o}pez}}, \citenamefont {{Nicodemi}},
  \citenamefont {{Pinton}},\ and\ \citenamefont {{Sellitto}}}]{BHP_2}%
  \BibitemOpen
  \bibfield  {author} {\bibinfo {author} {\bibfnamefont {S.~T.}\ \bibnamefont
  {{Bramwell}}}, \bibinfo {author} {\bibfnamefont {K.}~\bibnamefont
  {{Christensen}}}, \bibinfo {author} {\bibfnamefont {J.~Y.}\ \bibnamefont
  {{Fortin}}}, \bibinfo {author} {\bibfnamefont {P.~C.~W.}\ \bibnamefont
  {{Holdsworth}}}, \bibinfo {author} {\bibfnamefont {H.~J.}\ \bibnamefont
  {{Jensen}}}, \bibinfo {author} {\bibfnamefont {S.}~\bibnamefont {{Lise}}},
  \bibinfo {author} {\bibfnamefont {J.~M.}\ \bibnamefont {{L{\'o}pez}}},
  \bibinfo {author} {\bibfnamefont {M.}~\bibnamefont {{Nicodemi}}}, \bibinfo
  {author} {\bibfnamefont {J.~F.}\ \bibnamefont {{Pinton}}}, \ and\ \bibinfo
  {author} {\bibfnamefont {M.}~\bibnamefont {{Sellitto}}},\ }\href {\doibase
  10.1103/PhysRevLett.84.3744} {\bibfield  {journal} {\bibinfo  {journal}
  {\prl}\ }\textbf {\bibinfo {volume} {84}},\ \bibinfo {pages} {3744} (\bibinfo
  {year} {2000})}\BibitemShut {NoStop}%
\bibitem [{\citenamefont {{Portelli}}\ and\ \citenamefont
  {{Holdsworth}}(2002)}]{Portelli_02}%
  \BibitemOpen
  \bibfield  {author} {\bibinfo {author} {\bibfnamefont {B.}~\bibnamefont
  {{Portelli}}}\ and\ \bibinfo {author} {\bibfnamefont {P.~C.~W.}\ \bibnamefont
  {{Holdsworth}}},\ }\href {\doibase 10.1088/0305-4470/35/5/307} {\bibfield
  {journal} {\bibinfo  {journal} {Journal of Physics A Mathematical General}\
  }\textbf {\bibinfo {volume} {35}},\ \bibinfo {pages} {1231} (\bibinfo {year}
  {2002})}\BibitemShut {NoStop}%
\bibitem [{\citenamefont {Takeuchi}\ and\ \citenamefont
  {Sano}(2010)}]{Takeuchi2010}%
  \BibitemOpen
  \bibfield  {author} {\bibinfo {author} {\bibfnamefont {K.~A.}\ \bibnamefont
  {Takeuchi}}\ and\ \bibinfo {author} {\bibfnamefont {M.}~\bibnamefont
  {Sano}},\ }\href {\doibase 10.1103/PhysRevLett.104.230601} {\bibfield
  {journal} {\bibinfo  {journal} {Phys. Rev. Lett.}\ }\textbf {\bibinfo
  {volume} {104}},\ \bibinfo {pages} {230601} (\bibinfo {year}
  {2010})}\BibitemShut {NoStop}%
\bibitem [{\citenamefont {Martins}(2012)}]{Martins2012}%
  \BibitemOpen
  \bibfield  {author} {\bibinfo {author} {\bibfnamefont {P.~H.~L.}\
  \bibnamefont {Martins}},\ }\href {\doibase 10.1103/PhysRevE.85.041110}
  {\bibfield  {journal} {\bibinfo  {journal} {Phys. Rev. E}\ }\textbf {\bibinfo
  {volume} {85}},\ \bibinfo {pages} {041110} (\bibinfo {year}
  {2012})}\BibitemShut {NoStop}%
\bibitem [{\citenamefont {Malakis}\ \emph {et~al.}(2014)\citenamefont
  {Malakis}, \citenamefont {Fytas},\ and\ \citenamefont
  {G\"ulpinar}}]{Malakis2014}%
  \BibitemOpen
  \bibfield  {author} {\bibinfo {author} {\bibfnamefont {A.}~\bibnamefont
  {Malakis}}, \bibinfo {author} {\bibfnamefont {N.~G.}\ \bibnamefont {Fytas}},
  \ and\ \bibinfo {author} {\bibfnamefont {G.}~\bibnamefont {G\"ulpinar}},\
  }\href {\doibase 10.1103/PhysRevE.89.042103} {\bibfield  {journal} {\bibinfo
  {journal} {Phys. Rev. E}\ }\textbf {\bibinfo {volume} {89}},\ \bibinfo
  {pages} {042103} (\bibinfo {year} {2014})}\BibitemShut {NoStop}%
\bibitem [{\citenamefont {Xu}\ \emph {et~al.}(2020)\citenamefont {Xu},
  \citenamefont {Ferrenberg},\ and\ \citenamefont {Landau}}]{Xu2020}%
  \BibitemOpen
  \bibfield  {author} {\bibinfo {author} {\bibfnamefont {J.}~\bibnamefont
  {Xu}}, \bibinfo {author} {\bibfnamefont {A.~M.}\ \bibnamefont {Ferrenberg}},
  \ and\ \bibinfo {author} {\bibfnamefont {D.~P.}\ \bibnamefont {Landau}},\
  }\href {\doibase 10.1103/PhysRevE.101.023315} {\bibfield  {journal} {\bibinfo
   {journal} {Phys. Rev. E}\ }\textbf {\bibinfo {volume} {101}},\ \bibinfo
  {pages} {023315} (\bibinfo {year} {2020})}\BibitemShut {NoStop}%
\bibitem [{\citenamefont {Dyson}(1969)}]{Dyson1969}%
  \BibitemOpen
  \bibfield  {author} {\bibinfo {author} {\bibfnamefont {F.~J.}\ \bibnamefont
  {Dyson}},\ }\href {\doibase 10.1007/BF01645907} {\bibfield  {journal}
  {\bibinfo  {journal} {Communications in Mathematical Physics}\ }\textbf
  {\bibinfo {volume} {12}},\ \bibinfo {pages} {91} (\bibinfo {year}
  {1969})}\BibitemShut {NoStop}%
\bibitem [{\citenamefont {Bleher}\ and\ \citenamefont
  {Sinai}(1973)}]{Bleher1973}%
  \BibitemOpen
  \bibfield  {author} {\bibinfo {author} {\bibfnamefont {P.~M.}\ \bibnamefont
  {Bleher}}\ and\ \bibinfo {author} {\bibfnamefont {J.~G.}\ \bibnamefont
  {Sinai}},\ }\href {\doibase 10.1007/BF01645604} {\bibfield  {journal}
  {\bibinfo  {journal} {Communications in Mathematical Physics}\ }\textbf
  {\bibinfo {volume} {33}},\ \bibinfo {pages} {23} (\bibinfo {year}
  {1973})}\BibitemShut {NoStop}%
\bibitem [{\citenamefont {Collet}\ and\ \citenamefont
  {Eckmann}(1978)}]{ColletBook}%
  \BibitemOpen
  \bibfield  {author} {\bibinfo {author} {\bibfnamefont {P.}~\bibnamefont
  {Collet}}\ and\ \bibinfo {author} {\bibfnamefont {J.-P.}\ \bibnamefont
  {Eckmann}},\ }\href@noop {} {\emph {\bibinfo {title} {A Renormalization Group
  Analysis of the Hierarchical Model in Statistical Mechanics}}}\ (\bibinfo
  {publisher} {Springer Berlin, Heidelberg},\ \bibinfo {year}
  {1978})\BibitemShut {NoStop}%
\bibitem [{\citenamefont {Bleher}\ and\ \citenamefont
  {Major}(1987)}]{Bleher1987}%
  \BibitemOpen
  \bibfield  {author} {\bibinfo {author} {\bibfnamefont {P.~M.}\ \bibnamefont
  {Bleher}}\ and\ \bibinfo {author} {\bibfnamefont {P.}~\bibnamefont {Major}},\
  }\href {http://www.jstor.org/stable/2244058} {\bibfield  {journal} {\bibinfo
  {journal} {The Annals of Probability}\ }\textbf {\bibinfo {volume} {15}},\
  \bibinfo {pages} {431} (\bibinfo {year} {1987})}\BibitemShut {NoStop}%
\bibitem [{\citenamefont {Brankov}\ and\ \citenamefont
  {Danchev}(1989)}]{Brankov1989}%
  \BibitemOpen
  \bibfield  {author} {\bibinfo {author} {\bibfnamefont {J.}~\bibnamefont
  {Brankov}}\ and\ \bibinfo {author} {\bibfnamefont {D.}~\bibnamefont
  {Danchev}},\ }\href {\doibase https://doi.org/10.1016/0378-4371(89)90494-9}
  {\bibfield  {journal} {\bibinfo  {journal} {Physica A: Statistical Mechanics
  and its Applications}\ }\textbf {\bibinfo {volume} {158}},\ \bibinfo {pages}
  {842} (\bibinfo {year} {1989})}\BibitemShut {NoStop}%
\bibitem [{\citenamefont {Antal}\ \emph {et~al.}(2004)\citenamefont {Antal},
  \citenamefont {Droz},\ and\ \citenamefont {R{\'{a}}cz}}]{Antal2004}%
  \BibitemOpen
  \bibfield  {author} {\bibinfo {author} {\bibfnamefont {T.}~\bibnamefont
  {Antal}}, \bibinfo {author} {\bibfnamefont {M.}~\bibnamefont {Droz}}, \ and\
  \bibinfo {author} {\bibfnamefont {Z.}~\bibnamefont {R{\'{a}}cz}},\ }\href
  {\doibase 10.1088/0305-4470/37/5/001} {\bibfield  {journal} {\bibinfo
  {journal} {Journal of Physics A: Mathematical and General}\ }\textbf
  {\bibinfo {volume} {37}},\ \bibinfo {pages} {1465} (\bibinfo {year}
  {2004})}\BibitemShut {NoStop}%
\bibitem [{\citenamefont {Sasamoto}\ and\ \citenamefont
  {Spohn}(2010)}]{Sasamoto2010}%
  \BibitemOpen
  \bibfield  {author} {\bibinfo {author} {\bibfnamefont {T.}~\bibnamefont
  {Sasamoto}}\ and\ \bibinfo {author} {\bibfnamefont {H.}~\bibnamefont
  {Spohn}},\ }\href {\doibase 10.1103/PhysRevLett.104.230602} {\bibfield
  {journal} {\bibinfo  {journal} {Phys. Rev. Lett.}\ }\textbf {\bibinfo
  {volume} {104}},\ \bibinfo {pages} {230602} (\bibinfo {year}
  {2010})}\BibitemShut {NoStop}%
\bibitem [{\citenamefont {McCoy}\ and\ \citenamefont {Wu}(2013)}]{WuBook}%
  \BibitemOpen
  \bibfield  {author} {\bibinfo {author} {\bibfnamefont {B.~M.}\ \bibnamefont
  {McCoy}}\ and\ \bibinfo {author} {\bibfnamefont {T.~T.}\ \bibnamefont {Wu}},\
  }\href {\doibase doi:10.4159/harvard.9780674180758} {\emph {\bibinfo {title}
  {The Two-Dimensional Ising Model}}}\ (\bibinfo  {publisher} {Harvard
  University Press},\ \bibinfo {year} {2013})\BibitemShut {NoStop}%
\bibitem [{\citenamefont {Camia}\ \emph {et~al.}(2016)\citenamefont {Camia},
  \citenamefont {Garban},\ and\ \citenamefont {Newman}}]{Camia16}%
  \BibitemOpen
  \bibfield  {author} {\bibinfo {author} {\bibfnamefont {F.}~\bibnamefont
  {Camia}}, \bibinfo {author} {\bibfnamefont {C.}~\bibnamefont {Garban}}, \
  and\ \bibinfo {author} {\bibfnamefont {C.~M.}\ \bibnamefont {Newman}},\
  }\href {\doibase 10.1214/14-AIHP643} {\bibfield  {journal} {\bibinfo
  {journal} {Annales de l'Institut Henri Poincaré, Probabilités et
  Statistiques}\ }\textbf {\bibinfo {volume} {52}},\ \bibinfo {pages} {146 }
  (\bibinfo {year} {2016})}\BibitemShut {NoStop}%
\bibitem [{\citenamefont {Jona-Lasinio}(1975)}]{JonaLasino1975}%
  \BibitemOpen
  \bibfield  {author} {\bibinfo {author} {\bibfnamefont {G.}~\bibnamefont
  {Jona-Lasinio}},\ }\href@noop {} {\bibfield  {journal} {\bibinfo  {journal}
  {Il Nuovo Cimento B (1971-1996)}\ }\textbf {\bibinfo {volume} {26}},\
  \bibinfo {pages} {99} (\bibinfo {year} {1975})}\BibitemShut {NoStop}%
\bibitem [{\citenamefont {Gallavotti}\ and\ \citenamefont
  {Martin-L{\"o}f}(1975)}]{Gallavotti1975}%
  \BibitemOpen
  \bibfield  {author} {\bibinfo {author} {\bibfnamefont {G.}~\bibnamefont
  {Gallavotti}}\ and\ \bibinfo {author} {\bibfnamefont {A.}~\bibnamefont
  {Martin-L{\"o}f}},\ }\href@noop {} {\bibfield  {journal} {\bibinfo  {journal}
  {Il Nuovo Cimento B (1971-1996)}\ }\textbf {\bibinfo {volume} {25}},\
  \bibinfo {pages} {425} (\bibinfo {year} {1975})}\BibitemShut {NoStop}%
\bibitem [{\citenamefont {Cassandro}\ and\ \citenamefont
  {Jona-Lasinio}(1978)}]{Cassandro1978}%
  \BibitemOpen
  \bibfield  {author} {\bibinfo {author} {\bibfnamefont {M.}~\bibnamefont
  {Cassandro}}\ and\ \bibinfo {author} {\bibfnamefont {G.}~\bibnamefont
  {Jona-Lasinio}},\ }\href {\doibase 10.1080/00018737800101504} {\bibfield
  {journal} {\bibinfo  {journal} {Advances in Physics}\ }\textbf {\bibinfo
  {volume} {27}},\ \bibinfo {pages} {913} (\bibinfo {year} {1978})}\BibitemShut
  {NoStop}%
\bibitem [{\citenamefont {Wilson}\ and\ \citenamefont
  {Kogut}(1974)}]{WilsonKogut_74}%
  \BibitemOpen
  \bibfield  {author} {\bibinfo {author} {\bibfnamefont {K.~G.}\ \bibnamefont
  {Wilson}}\ and\ \bibinfo {author} {\bibfnamefont {J.~B.}\ \bibnamefont
  {Kogut}},\ }\href {\doibase 10.1016/0370-1573(74)90023-4} {\bibfield
  {journal} {\bibinfo  {journal} {Phys. Rept.}\ }\textbf {\bibinfo {volume}
  {12}},\ \bibinfo {pages} {75} (\bibinfo {year} {1974})}\BibitemShut {NoStop}%
\bibitem [{\citenamefont {Bruce}\ \emph {et~al.}(1979)\citenamefont {Bruce},
  \citenamefont {Schneider},\ and\ \citenamefont {Stoll}}]{Bruce1979}%
  \BibitemOpen
  \bibfield  {author} {\bibinfo {author} {\bibfnamefont {A.~D.}\ \bibnamefont
  {Bruce}}, \bibinfo {author} {\bibfnamefont {T.}~\bibnamefont {Schneider}}, \
  and\ \bibinfo {author} {\bibfnamefont {E.}~\bibnamefont {Stoll}},\ }\href
  {\doibase 10.1103/PhysRevLett.43.1284} {\bibfield  {journal} {\bibinfo
  {journal} {Phys. Rev. Lett.}\ }\textbf {\bibinfo {volume} {43}},\ \bibinfo
  {pages} {1284} (\bibinfo {year} {1979})}\BibitemShut {NoStop}%
\bibitem [{\citenamefont {Eisenriegler}\ and\ \citenamefont
  {Tomaschitz}(1987)}]{Eisenriegler1987}%
  \BibitemOpen
  \bibfield  {author} {\bibinfo {author} {\bibfnamefont {E.}~\bibnamefont
  {Eisenriegler}}\ and\ \bibinfo {author} {\bibfnamefont {R.}~\bibnamefont
  {Tomaschitz}},\ }\href {\doibase 10.1103/PhysRevB.35.4876} {\bibfield
  {journal} {\bibinfo  {journal} {Phys. Rev. B}\ }\textbf {\bibinfo {volume}
  {35}},\ \bibinfo {pages} {4876} (\bibinfo {year} {1987})}\BibitemShut
  {NoStop}%
\bibitem [{\citenamefont {Hilfer}(1993)}]{Hilfer1993}%
  \BibitemOpen
  \bibfield  {author} {\bibinfo {author} {\bibfnamefont {R.}~\bibnamefont
  {Hilfer}},\ }\href {\doibase 10.1142/S0217979293003711} {\bibfield  {journal}
  {\bibinfo  {journal} {International Journal of Modern Physics B}\ }\textbf
  {\bibinfo {volume} {07}},\ \bibinfo {pages} {4371} (\bibinfo {year}
  {1993})}\BibitemShut {NoStop}%
\bibitem [{\citenamefont {Hilfer}\ and\ \citenamefont
  {Wilding}(1995)}]{Hilfer1995}%
  \BibitemOpen
  \bibfield  {author} {\bibinfo {author} {\bibfnamefont {R.}~\bibnamefont
  {Hilfer}}\ and\ \bibinfo {author} {\bibfnamefont {N.~B.}\ \bibnamefont
  {Wilding}},\ }\href {\doibase 10.1088/0305-4470/28/10/001} {\bibfield
  {journal} {\bibinfo  {journal} {Journal of Physics A: Mathematical and
  General}\ }\textbf {\bibinfo {volume} {28}},\ \bibinfo {pages} {L281}
  (\bibinfo {year} {1995})}\BibitemShut {NoStop}%
\bibitem [{\citenamefont {Esser}\ \emph {et~al.}(1995)\citenamefont {Esser},
  \citenamefont {Dohm},\ and\ \citenamefont {Chen}}]{Esser1995}%
  \BibitemOpen
  \bibfield  {author} {\bibinfo {author} {\bibfnamefont {A.}~\bibnamefont
  {Esser}}, \bibinfo {author} {\bibfnamefont {V.}~\bibnamefont {Dohm}}, \ and\
  \bibinfo {author} {\bibfnamefont {X.}~\bibnamefont {Chen}},\ }\href {\doibase
  https://doi.org/10.1016/0378-4371(95)00264-2} {\bibfield  {journal} {\bibinfo
   {journal} {Physica A: Statistical Mechanics and its Applications}\ }\textbf
  {\bibinfo {volume} {222}},\ \bibinfo {pages} {355} (\bibinfo {year}
  {1995})}\BibitemShut {NoStop}%
\bibitem [{\citenamefont {Bruce}(1997)}]{Bruce1997}%
  \BibitemOpen
  \bibfield  {author} {\bibinfo {author} {\bibfnamefont {A.~D.}\ \bibnamefont
  {Bruce}},\ }\href {\doibase 10.1103/PhysRevE.55.2315} {\bibfield  {journal}
  {\bibinfo  {journal} {Phys. Rev. E}\ }\textbf {\bibinfo {volume} {55}},\
  \bibinfo {pages} {2315} (\bibinfo {year} {1997})}\BibitemShut {NoStop}%
\bibitem [{\citenamefont {Rudnick}\ \emph {et~al.}(1998)\citenamefont
  {Rudnick}, \citenamefont {Lay},\ and\ \citenamefont {Jasnow}}]{Rudnick1998}%
  \BibitemOpen
  \bibfield  {author} {\bibinfo {author} {\bibfnamefont {J.}~\bibnamefont
  {Rudnick}}, \bibinfo {author} {\bibfnamefont {W.}~\bibnamefont {Lay}}, \ and\
  \bibinfo {author} {\bibfnamefont {D.}~\bibnamefont {Jasnow}},\ }\href
  {\doibase 10.1103/PhysRevE.58.2902} {\bibfield  {journal} {\bibinfo
  {journal} {Phys. Rev. E}\ }\textbf {\bibinfo {volume} {58}},\ \bibinfo
  {pages} {2902} (\bibinfo {year} {1998})}\BibitemShut {NoStop}%
\bibitem [{\citenamefont {Berges}\ \emph {et~al.}(2002)\citenamefont {Berges},
  \citenamefont {Tetradis},\ and\ \citenamefont {Wetterich}}]{Berges_02}%
  \BibitemOpen
  \bibfield  {author} {\bibinfo {author} {\bibfnamefont {J.}~\bibnamefont
  {Berges}}, \bibinfo {author} {\bibfnamefont {N.}~\bibnamefont {Tetradis}}, \
  and\ \bibinfo {author} {\bibfnamefont {C.}~\bibnamefont {Wetterich}},\ }\href
  {\doibase DOI: 10.1016/S0370-1573(01)00098-9} {\bibfield  {journal} {\bibinfo
   {journal} {Physics Reports}\ }\textbf {\bibinfo {volume} {363}},\ \bibinfo
  {pages} {223 } (\bibinfo {year} {2002})}\BibitemShut {NoStop}%
\bibitem [{\citenamefont {Dupuis}\ \emph {et~al.}(2021)\citenamefont {Dupuis},
  \citenamefont {Canet}, \citenamefont {Eichhorn}, \citenamefont {Metzner},
  \citenamefont {Pawlowski}, \citenamefont {Tissier},\ and\ \citenamefont
  {Wschebor}}]{Dupuis2021}%
  \BibitemOpen
  \bibfield  {author} {\bibinfo {author} {\bibfnamefont {N.}~\bibnamefont
  {Dupuis}}, \bibinfo {author} {\bibfnamefont {L.}~\bibnamefont {Canet}},
  \bibinfo {author} {\bibfnamefont {A.}~\bibnamefont {Eichhorn}}, \bibinfo
  {author} {\bibfnamefont {W.}~\bibnamefont {Metzner}}, \bibinfo {author}
  {\bibfnamefont {J.}~\bibnamefont {Pawlowski}}, \bibinfo {author}
  {\bibfnamefont {M.}~\bibnamefont {Tissier}}, \ and\ \bibinfo {author}
  {\bibfnamefont {N.}~\bibnamefont {Wschebor}},\ }\href {\doibase
  https://doi.org/10.1016/j.physrep.2021.01.001} {\bibfield  {journal}
  {\bibinfo  {journal} {Physics Reports}\ }\textbf {\bibinfo {volume} {910}},\
  \bibinfo {pages} {1} (\bibinfo {year} {2021})}\BibitemShut {NoStop}%
\bibitem [{\citenamefont {Zinn-Justin}(2007)}]{ZinnForChildren}%
  \BibitemOpen
  \bibfield  {author} {\bibinfo {author} {\bibfnamefont {J.}~\bibnamefont
  {Zinn-Justin}},\ }\href@noop {} {\emph {\bibinfo {title} {Phase transitions
  and renormalization group}}}\ (\bibinfo {year} {2007})\BibitemShut {NoStop}%
\bibitem [{\citenamefont {Bouchaud}\ and\ \citenamefont
  {Georges}(1990)}]{Bouchaud1990}%
  \BibitemOpen
  \bibfield  {author} {\bibinfo {author} {\bibfnamefont {J.-P.}\ \bibnamefont
  {Bouchaud}}\ and\ \bibinfo {author} {\bibfnamefont {A.}~\bibnamefont
  {Georges}},\ }\href {\doibase https://doi.org/10.1016/0370-1573(90)90099-N}
  {\bibfield  {journal} {\bibinfo  {journal} {Physics Reports}\ }\textbf
  {\bibinfo {volume} {195}},\ \bibinfo {pages} {127} (\bibinfo {year}
  {1990})}\BibitemShut {NoStop}%
\bibitem [{Note1()}]{Note1}%
  \BibitemOpen
  \bibinfo {note} {The shapes of the family of PDFs also depend crucially on
  the boundary conditions \cite {Binder1981a}. Here we focus on periodic
  boundary conditions only.}\BibitemShut {Stop}%
\bibitem [{\citenamefont {Fukuda}\ and\ \citenamefont
  {Kyriakopoulos}(1975)}]{Fukuda1975}%
  \BibitemOpen
  \bibfield  {author} {\bibinfo {author} {\bibfnamefont {R.}~\bibnamefont
  {Fukuda}}\ and\ \bibinfo {author} {\bibfnamefont {E.}~\bibnamefont
  {Kyriakopoulos}},\ }\href {\doibase
  https://doi.org/10.1016/0550-3213(75)90014-0} {\bibfield  {journal} {\bibinfo
   {journal} {Nuclear Physics B}\ }\textbf {\bibinfo {volume} {85}},\ \bibinfo
  {pages} {354} (\bibinfo {year} {1975})}\BibitemShut {NoStop}%
\bibitem [{\citenamefont {O'Raifeartaigh}\ \emph {et~al.}(1986)\citenamefont
  {O'Raifeartaigh}, \citenamefont {Wipf},\ and\ \citenamefont
  {Yoneyama}}]{ORaifeartaigh1986}%
  \BibitemOpen
  \bibfield  {author} {\bibinfo {author} {\bibfnamefont {L.}~\bibnamefont
  {O'Raifeartaigh}}, \bibinfo {author} {\bibfnamefont {A.}~\bibnamefont
  {Wipf}}, \ and\ \bibinfo {author} {\bibfnamefont {H.}~\bibnamefont
  {Yoneyama}},\ }\href {\doibase https://doi.org/10.1016/S0550-3213(86)80031-1}
  {\bibfield  {journal} {\bibinfo  {journal} {Nuclear Physics B}\ }\textbf
  {\bibinfo {volume} {271}},\ \bibinfo {pages} {653} (\bibinfo {year}
  {1986})}\BibitemShut {NoStop}%
\bibitem [{\citenamefont {G\"ockeler}\ and\ \citenamefont
  {Leutwyler}(1991)}]{Gockeler1991}%
  \BibitemOpen
  \bibfield  {author} {\bibinfo {author} {\bibfnamefont {M.}~\bibnamefont
  {G\"ockeler}}\ and\ \bibinfo {author} {\bibfnamefont {H.}~\bibnamefont
  {Leutwyler}},\ }\href {\doibase https://doi.org/10.1016/0550-3213(91)90260-5}
  {\bibfield  {journal} {\bibinfo  {journal} {Nuclear Physics B}\ }\textbf
  {\bibinfo {volume} {350}},\ \bibinfo {pages} {228} (\bibinfo {year}
  {1991})}\BibitemShut {NoStop}%
\bibitem [{\citenamefont {Wetterich}(1991)}]{Wetterich1991}%
  \BibitemOpen
  \bibfield  {author} {\bibinfo {author} {\bibfnamefont {C.}~\bibnamefont
  {Wetterich}},\ }\href {\doibase DOI: 10.1016/0550-3213(91)90099-J} {\bibfield
   {journal} {\bibinfo  {journal} {Nuclear Physics B}\ }\textbf {\bibinfo
  {volume} {352}},\ \bibinfo {pages} {529 } (\bibinfo {year}
  {1991})}\BibitemShut {NoStop}%
\bibitem [{\citenamefont {Wetterich}(1993{\natexlab{a}})}]{Wetterich1993}%
  \BibitemOpen
  \bibfield  {author} {\bibinfo {author} {\bibfnamefont {C.}~\bibnamefont
  {Wetterich}},\ }\href {\doibase DOI: 10.1016/0370-2693(93)90726-X} {\bibfield
   {journal} {\bibinfo  {journal} {Physics Letters B}\ }\textbf {\bibinfo
  {volume} {301}},\ \bibinfo {pages} {90 } (\bibinfo {year}
  {1993}{\natexlab{a}})}\BibitemShut {NoStop}%
\bibitem [{\citenamefont {Wetterich}(1993{\natexlab{b}})}]{Wetterich1993a}%
  \BibitemOpen
  \bibfield  {author} {\bibinfo {author} {\bibfnamefont {C.}~\bibnamefont
  {Wetterich}},\ }\href {http://dx.doi.org/10.1007/BF01560044} {\bibfield
  {journal} {\bibinfo  {journal} {Zeitschrift f\"ur Physik C}\ }\textbf
  {\bibinfo {volume} {60}},\ \bibinfo {pages} {461} (\bibinfo {year}
  {1993}{\natexlab{b}})}\BibitemShut {NoStop}%
\bibitem [{\citenamefont {Canet}\ \emph {et~al.}(2003)\citenamefont {Canet},
  \citenamefont {Delamotte}, \citenamefont {Mouhanna},\ and\ \citenamefont
  {Vidal}}]{PhysRevB.68.064421}%
  \BibitemOpen
  \bibfield  {author} {\bibinfo {author} {\bibfnamefont {L.}~\bibnamefont
  {Canet}}, \bibinfo {author} {\bibfnamefont {B.}~\bibnamefont {Delamotte}},
  \bibinfo {author} {\bibfnamefont {D.}~\bibnamefont {Mouhanna}}, \ and\
  \bibinfo {author} {\bibfnamefont {J.}~\bibnamefont {Vidal}},\ }\href
  {\doibase 10.1103/PhysRevB.68.064421} {\bibfield  {journal} {\bibinfo
  {journal} {Phys. Rev. B}\ }\textbf {\bibinfo {volume} {68}},\ \bibinfo
  {pages} {064421} (\bibinfo {year} {2003})}\BibitemShut {NoStop}%
\bibitem [{\citenamefont {Balog}\ \emph {et~al.}(2019)\citenamefont {Balog},
  \citenamefont {Chat{\'e}}, \citenamefont {Delamotte}, \citenamefont
  {Marohni{\'c}},\ and\ \citenamefont {Wschebor}}]{Balog2019}%
  \BibitemOpen
  \bibfield  {author} {\bibinfo {author} {\bibfnamefont {I.}~\bibnamefont
  {Balog}}, \bibinfo {author} {\bibfnamefont {H.}~\bibnamefont {Chat{\'e}}},
  \bibinfo {author} {\bibfnamefont {B.}~\bibnamefont {Delamotte}}, \bibinfo
  {author} {\bibfnamefont {M.}~\bibnamefont {Marohni{\'c}}}, \ and\ \bibinfo
  {author} {\bibfnamefont {N.}~\bibnamefont {Wschebor}},\ }\href {\doibase
  10.1103/PhysRevLett.123.240604} {\bibfield  {journal} {\bibinfo  {journal}
  {Physical Review Letters}\ }\textbf {\bibinfo {volume} {123}},\ \bibinfo
  {pages} {240604} (\bibinfo {year} {2019})}\BibitemShut {NoStop}%
\bibitem [{\citenamefont {De~Polsi}\ \emph {et~al.}(2020)\citenamefont
  {De~Polsi}, \citenamefont {Balog}, \citenamefont {Tissier},\ and\
  \citenamefont {Wschebor}}]{PhysRevE.101.042113}%
  \BibitemOpen
  \bibfield  {author} {\bibinfo {author} {\bibfnamefont {G.}~\bibnamefont
  {De~Polsi}}, \bibinfo {author} {\bibfnamefont {I.}~\bibnamefont {Balog}},
  \bibinfo {author} {\bibfnamefont {M.}~\bibnamefont {Tissier}}, \ and\
  \bibinfo {author} {\bibfnamefont {N.}~\bibnamefont {Wschebor}},\ }\href
  {\doibase 10.1103/PhysRevE.101.042113} {\bibfield  {journal} {\bibinfo
  {journal} {Phys. Rev. E}\ }\textbf {\bibinfo {volume} {101}},\ \bibinfo
  {pages} {042113} (\bibinfo {year} {2020})}\BibitemShut {NoStop}%
\bibitem [{SM()}]{SM}%
  \BibitemOpen
  \href@noop {} {}\bibinfo {note} {See Supplementary Material the derivation of
  $\Gamma_{M,k}$, its flow equation, and the numerical solution of the flow;
  details on the Monte-Carlo simulations. It also includes Refs. [56-59].}\BibitemShut {Stop}%
\bibitem [{\citenamefont {Fister}\ and\ \citenamefont
  {Pawlowski}(2015)}]{Fister2015}%
  \BibitemOpen
  \bibfield  {author} {\bibinfo {author} {\bibfnamefont {L.}~\bibnamefont
  {Fister}}\ and\ \bibinfo {author} {\bibfnamefont {J.~M.}\ \bibnamefont
  {Pawlowski}},\ }\href {\doibase 10.1103/PhysRevD.92.076009} {\bibfield
  {journal} {\bibinfo  {journal} {Phys. Rev. D}\ }\textbf {\bibinfo {volume}
  {92}},\ \bibinfo {pages} {076009} (\bibinfo {year} {2015})}\BibitemShut
  {NoStop}%
  \bibitem [{\citenamefont {Lundow}\ and\ \citenamefont
  {Campbell}(2018)}]{Lundow2018}%
  \BibitemOpen
  \bibfield  {author} {\bibinfo {author} {\bibfnamefont {P.}~\bibnamefont
  {Lundow}}\ and\ \bibinfo {author} {\bibfnamefont {I.}~\bibnamefont
  {Campbell}},\ }\href {\doibase https://doi.org/10.1016/j.physa.2018.06.087}
  {\bibfield  {journal} {\bibinfo  {journal} {Physica A}\ }\textbf {\bibinfo
  {volume} {511}},\ \bibinfo {pages} {40} (\bibinfo {year} {2018})}\BibitemShut
  {NoStop}%
\bibitem [{\citenamefont {Ferrenberg}\ \emph {et~al.}(2018)\citenamefont
  {Ferrenberg}, \citenamefont {Xu},\ and\ \citenamefont
  {Landau}}]{Ferrenberg2018}%
  \BibitemOpen
  \bibfield  {author} {\bibinfo {author} {\bibfnamefont {A.~M.}\ \bibnamefont
  {Ferrenberg}}, \bibinfo {author} {\bibfnamefont {J.}~\bibnamefont {Xu}}, \
  and\ \bibinfo {author} {\bibfnamefont {D.~P.}\ \bibnamefont {Landau}},\
  }\href {\doibase 10.1103/PhysRevE.97.043301} {\bibfield  {journal} {\bibinfo
  {journal} {Phys. Rev. E}\ }\textbf {\bibinfo {volume} {97}},\ \bibinfo
  {pages} {043301} (\bibinfo {year} {2018})}\BibitemShut {NoStop}%
\bibitem [{\citenamefont {Kos}\ \emph {et~al.}(2016)\citenamefont {Kos},
  \citenamefont {Poland}, \citenamefont {Simmons-Duffin},\ and\ \citenamefont
  {Vichi}}]{Kos2016}%
  \BibitemOpen
  \bibfield  {author} {\bibinfo {author} {\bibfnamefont {F.}~\bibnamefont
  {Kos}}, \bibinfo {author} {\bibfnamefont {D.}~\bibnamefont {Poland}},
  \bibinfo {author} {\bibfnamefont {D.}~\bibnamefont {Simmons-Duffin}}, \ and\
  \bibinfo {author} {\bibfnamefont {A.}~\bibnamefont {Vichi}},\ }\href
  {\doibase 10.1007/JHEP08(2016)036} {\bibfield  {journal} {\bibinfo  {journal}
  {Journal of High Energy Physics}\ }\textbf {\bibinfo {volume} {2016}},\
  \bibinfo {pages} {36} (\bibinfo {year} {2016})}\BibitemShut {NoStop}%
\bibitem [{\citenamefont {Wolff}(1989)}]{Wolff1989}%
  \BibitemOpen
  \bibfield  {author} {\bibinfo {author} {\bibfnamefont {U.}~\bibnamefont
  {Wolff}},\ }\href {\doibase 10.1103/PhysRevLett.62.361} {\bibfield  {journal}
  {\bibinfo  {journal} {Phys. Rev. Lett.}\ }\textbf {\bibinfo {volume} {62}},\
  \bibinfo {pages} {361} (\bibinfo {year} {1989})}\BibitemShut {NoStop}%
\bibitem [{\citenamefont {Ferrenberg}\ and\ \citenamefont
  {Swendsen}(1988)}]{Ferrenberg1988}%
  \BibitemOpen
  \bibfield  {author} {\bibinfo {author} {\bibfnamefont {A.~M.}\ \bibnamefont
  {Ferrenberg}}\ and\ \bibinfo {author} {\bibfnamefont {R.~H.}\ \bibnamefont
  {Swendsen}},\ }\href {\doibase 10.1103/PhysRevLett.61.2635} {\bibfield
  {journal} {\bibinfo  {journal} {Phys. Rev. Lett.}\ }\textbf {\bibinfo
  {volume} {61}},\ \bibinfo {pages} {2635} (\bibinfo {year}
  {1988})}\BibitemShut {NoStop}%
\bibitem [{nex()}]{next_paper_2035}%
  \BibitemOpen
  \href@noop {} {}\bibinfo {note} {I. Balog, A. Ran\c{c}on and B. Delamotte, to
  be published.}\BibitemShut {Stop}%
\bibitem [{\citenamefont {Tarjus}\ and\ \citenamefont
  {Tissier}(2004)}]{PhysRevLett.93.267008}%
  \BibitemOpen
  \bibfield  {author} {\bibinfo {author} {\bibfnamefont {G.}~\bibnamefont
  {Tarjus}}\ and\ \bibinfo {author} {\bibfnamefont {M.}~\bibnamefont
  {Tissier}},\ }\href {\doibase 10.1103/PhysRevLett.93.267008} {\bibfield
  {journal} {\bibinfo  {journal} {Phys. Rev. Lett.}\ }\textbf {\bibinfo
  {volume} {93}},\ \bibinfo {pages} {267008} (\bibinfo {year}
  {2004})}\BibitemShut {NoStop}%
\bibitem [{\citenamefont {Tissier}\ and\ \citenamefont
  {Tarjus}(2011)}]{PhysRevLett.107.041601}%
  \BibitemOpen
  \bibfield  {author} {\bibinfo {author} {\bibfnamefont {M.}~\bibnamefont
  {Tissier}}\ and\ \bibinfo {author} {\bibfnamefont {G.}~\bibnamefont
  {Tarjus}},\ }\href {\doibase 10.1103/PhysRevLett.107.041601} {\bibfield
  {journal} {\bibinfo  {journal} {Phys. Rev. Lett.}\ }\textbf {\bibinfo
  {volume} {107}},\ \bibinfo {pages} {041601} (\bibinfo {year}
  {2011})}\BibitemShut {NoStop}%
\bibitem [{\citenamefont {Tissier}\ and\ \citenamefont
  {Tarjus}(2012)}]{PhysRevB.85.104202}%
  \BibitemOpen
  \bibfield  {author} {\bibinfo {author} {\bibfnamefont {M.}~\bibnamefont
  {Tissier}}\ and\ \bibinfo {author} {\bibfnamefont {G.}~\bibnamefont
  {Tarjus}},\ }\href {\doibase 10.1103/PhysRevB.85.104202} {\bibfield
  {journal} {\bibinfo  {journal} {Phys. Rev. B}\ }\textbf {\bibinfo {volume}
  {85}},\ \bibinfo {pages} {104202} (\bibinfo {year} {2012})}\BibitemShut
  {NoStop}%
\bibitem [{\citenamefont {Canet}\ \emph {et~al.}(2004)\citenamefont {Canet},
  \citenamefont {Delamotte}, \citenamefont {Deloubri\`ere},\ and\ \citenamefont
  {Wschebor}}]{PhysRevLett.92.195703}%
  \BibitemOpen
  \bibfield  {author} {\bibinfo {author} {\bibfnamefont {L.}~\bibnamefont
  {Canet}}, \bibinfo {author} {\bibfnamefont {B.}~\bibnamefont {Delamotte}},
  \bibinfo {author} {\bibfnamefont {O.}~\bibnamefont {Deloubri\`ere}}, \ and\
  \bibinfo {author} {\bibfnamefont {N.}~\bibnamefont {Wschebor}},\ }\href
  {\doibase 10.1103/PhysRevLett.92.195703} {\bibfield  {journal} {\bibinfo
  {journal} {Phys. Rev. Lett.}\ }\textbf {\bibinfo {volume} {92}},\ \bibinfo
  {pages} {195703} (\bibinfo {year} {2004})}\BibitemShut {NoStop}%
\bibitem [{\citenamefont {Canet}\ \emph {et~al.}(2005)\citenamefont {Canet},
  \citenamefont {Chat\'e}, \citenamefont {Delamotte}, \citenamefont {Dornic},\
  and\ \citenamefont {Mu\~noz}}]{PhysRevLett.95.100601}%
  \BibitemOpen
  \bibfield  {author} {\bibinfo {author} {\bibfnamefont {L.}~\bibnamefont
  {Canet}}, \bibinfo {author} {\bibfnamefont {H.}~\bibnamefont {Chat\'e}},
  \bibinfo {author} {\bibfnamefont {B.}~\bibnamefont {Delamotte}}, \bibinfo
  {author} {\bibfnamefont {I.}~\bibnamefont {Dornic}}, \ and\ \bibinfo {author}
  {\bibfnamefont {M.~A.}\ \bibnamefont {Mu\~noz}},\ }\href {\doibase
  10.1103/PhysRevLett.95.100601} {\bibfield  {journal} {\bibinfo  {journal}
  {Phys. Rev. Lett.}\ }\textbf {\bibinfo {volume} {95}},\ \bibinfo {pages}
  {100601} (\bibinfo {year} {2005})}\BibitemShut {NoStop}%
\bibitem [{\citenamefont {Canet}\ \emph {et~al.}(2010)\citenamefont {Canet},
  \citenamefont {Chat\'e}, \citenamefont {Delamotte},\ and\ \citenamefont
  {Wschebor}}]{PhysRevLett.104.150601}%
  \BibitemOpen
  \bibfield  {author} {\bibinfo {author} {\bibfnamefont {L.}~\bibnamefont
  {Canet}}, \bibinfo {author} {\bibfnamefont {H.}~\bibnamefont {Chat\'e}},
  \bibinfo {author} {\bibfnamefont {B.}~\bibnamefont {Delamotte}}, \ and\
  \bibinfo {author} {\bibfnamefont {N.}~\bibnamefont {Wschebor}},\ }\href
  {\doibase 10.1103/PhysRevLett.104.150601} {\bibfield  {journal} {\bibinfo
  {journal} {Phys. Rev. Lett.}\ }\textbf {\bibinfo {volume} {104}},\ \bibinfo
  {pages} {150601} (\bibinfo {year} {2010})}\BibitemShut {NoStop}%
\bibitem [{\citenamefont {Canet}\ \emph {et~al.}(2011)\citenamefont {Canet},
  \citenamefont {Chat\'e}, \citenamefont {Delamotte},\ and\ \citenamefont
  {Wschebor}}]{PhysRevE.84.061128}%
  \BibitemOpen
  \bibfield  {author} {\bibinfo {author} {\bibfnamefont {L.}~\bibnamefont
  {Canet}}, \bibinfo {author} {\bibfnamefont {H.}~\bibnamefont {Chat\'e}},
  \bibinfo {author} {\bibfnamefont {B.}~\bibnamefont {Delamotte}}, \ and\
  \bibinfo {author} {\bibfnamefont {N.}~\bibnamefont {Wschebor}},\ }\href
  {\doibase 10.1103/PhysRevE.84.061128} {\bibfield  {journal} {\bibinfo
  {journal} {Phys. Rev. E}\ }\textbf {\bibinfo {volume} {84}},\ \bibinfo
  {pages} {061128} (\bibinfo {year} {2011})}\BibitemShut {NoStop}%
\bibitem [{\citenamefont {Kloss}\ \emph {et~al.}(2014)\citenamefont {Kloss},
  \citenamefont {Canet}, \citenamefont {Delamotte},\ and\ \citenamefont
  {Wschebor}}]{PhysRevE.89.022108}%
  \BibitemOpen
  \bibfield  {author} {\bibinfo {author} {\bibfnamefont {T.}~\bibnamefont
  {Kloss}}, \bibinfo {author} {\bibfnamefont {L.}~\bibnamefont {Canet}},
  \bibinfo {author} {\bibfnamefont {B.}~\bibnamefont {Delamotte}}, \ and\
  \bibinfo {author} {\bibfnamefont {N.}~\bibnamefont {Wschebor}},\ }\href
  {\doibase 10.1103/PhysRevE.89.022108} {\bibfield  {journal} {\bibinfo
  {journal} {Phys. Rev. E}\ }\textbf {\bibinfo {volume} {89}},\ \bibinfo
  {pages} {022108} (\bibinfo {year} {2014})}\BibitemShut {NoStop}%
\bibitem [{\citenamefont {Canet}\ \emph {et~al.}(2016)\citenamefont {Canet},
  \citenamefont {Delamotte},\ and\ \citenamefont
  {Wschebor}}]{PhysRevE.93.063101}%
  \BibitemOpen
  \bibfield  {author} {\bibinfo {author} {\bibfnamefont {L.}~\bibnamefont
  {Canet}}, \bibinfo {author} {\bibfnamefont {B.}~\bibnamefont {Delamotte}}, \
  and\ \bibinfo {author} {\bibfnamefont {N.}~\bibnamefont {Wschebor}},\ }\href
  {\doibase 10.1103/PhysRevE.93.063101} {\bibfield  {journal} {\bibinfo
  {journal} {Phys. Rev. E}\ }\textbf {\bibinfo {volume} {93}},\ \bibinfo
  {pages} {063101} (\bibinfo {year} {2016})}\BibitemShut {NoStop}%
\bibitem [{\citenamefont {Blaizot}\ \emph {et~al.}(2006)\citenamefont
  {Blaizot}, \citenamefont {M\'endez-Galain},\ and\ \citenamefont
  {Wschebor}}]{Blaizot2006}%
  \BibitemOpen
  \bibfield  {author} {\bibinfo {author} {\bibfnamefont {J.-P.}\ \bibnamefont
  {Blaizot}}, \bibinfo {author} {\bibfnamefont {R.}~\bibnamefont
  {M\'endez-Galain}}, \ and\ \bibinfo {author} {\bibfnamefont {N.}~\bibnamefont
  {Wschebor}},\ }\href
  {http://www.sciencedirect.com/science/article/B6TVN-4HHX2B1-3/2/0a319fce96e1979563ee56c3422f5477}
  {\bibfield  {journal} {\bibinfo  {journal} {Physics Letters B}\ }\textbf
  {\bibinfo {volume} {632}} (\bibinfo {year} {2006})}\BibitemShut {NoStop}%
\bibitem [{\citenamefont {Benitez}\ \emph {et~al.}(2009)\citenamefont
  {Benitez}, \citenamefont {Blaizot}, \citenamefont {Chat\'e}, \citenamefont
  {Delamotte}, \citenamefont {M\'endez-Galain},\ and\ \citenamefont
  {Wschebor}}]{Benitez2009}%
  \BibitemOpen
  \bibfield  {author} {\bibinfo {author} {\bibfnamefont {F.}~\bibnamefont
  {Benitez}}, \bibinfo {author} {\bibfnamefont {J.-P.}\ \bibnamefont
  {Blaizot}}, \bibinfo {author} {\bibfnamefont {H.}~\bibnamefont {Chat\'e}},
  \bibinfo {author} {\bibfnamefont {B.}~\bibnamefont {Delamotte}}, \bibinfo
  {author} {\bibfnamefont {R.}~\bibnamefont {M\'endez-Galain}}, \ and\ \bibinfo
  {author} {\bibfnamefont {N.}~\bibnamefont {Wschebor}},\ }\href {\doibase
  10.1103/PhysRevE.80.030103} {\bibfield  {journal} {\bibinfo  {journal} {Phys.
  Rev. E}\ }\textbf {\bibinfo {volume} {80}},\ \bibinfo {pages} {030103(R)}
  (\bibinfo {year} {2009})}\BibitemShut {NoStop}%
\end{thebibliography}%
\bibliographystyle{apsrev4-1} 

\section*{Supplemental Materials}

\section{Definition of the effective actions $\Gamma_{M,k}$}

We start from the $M$-dependent partition functions in presence of a regulator $R_k$ and a source $h$ as defined in the main text:
\begin{equation}
\mathcal Z_{M,k}[h] =\int \DD\hat\phi\, e^{-\mathcal H_M[\hat\phi] + h.\hat\phi - \frac12 \hat\phi.R_k.\hat\phi}
\end{equation}
where ${\cal H}_M[\hat\phi]={\cal H}(\hat\phi)+\frac {M^2}2 [\int_\x (\hat\phi(\x)-s)]^2$. 

Using $\phi(\x)=\frac{\delta { \mathcal Z}_{M,k}}{\delta h(\x)}$, we define the modified Legendre transform
\begin{equation}
\Gamma_{M,k}[\phi]=-\ln \mathcal Z_{M,k}[h]+h.\phi-\frac12\phi. R_k.\phi-\frac {M^2}2 \left[\int_\x (\phi(\x)-s)\right]^2
\label{eq_def_GammaA_LT}
\end{equation}
Note that
\begin{equation}
\frac{\delta\Gamma_{M,k}}{\delta\phi(\x)}=h(\x)-\int_\y R_k(\x,\y)\phi(\y)-M^2\int_\y(\phi(\y)-s).
\end{equation}

which allows us to write
\begin{equation}
\begin{split}
e^{-\Gamma_{M,k}[\phi]}=&\int \DD\hat\phi\, e^{-{\cal H}(\hat\phi)+\frac{\delta\Gamma_{M,k}}{\delta\phi}.(\hat\phi-\phi) - \frac12 (\hat\phi-\phi).R_{k}.(\hat\phi-\phi)}\, \times\\
&\ \ \ \ \ \ \ \  e^{-\frac {M^2}2 [\int_\x (\hat\phi(\x)-\phi(\x))]^2}.
\end{split}
\end{equation}
Defining $R_{M,k}(\x,\y)=R_{k}(\x,\y)+M^2$, this equation can be rewritten as:
\begin{equation}
e^{-\Gamma_{M,k}[\phi]}=\int \DD\hat\phi\, e^{-{\cal H}(\hat\phi)+\frac{\delta\Gamma_{M,k}}{\delta\phi}.(\hat\phi-\phi) - \frac12 (\hat\phi-\phi).R_{M,k}.(\hat\phi-\phi) },
\label{eq_def_GammaA_LT_nice}
\end{equation}
and therefore $\Gamma_{M,k}[\phi]$ can be interpreted as the (modified) Gibbs free energy of a system of Hamiltonian $\cal H$ instead of ${\cal H}_M$ regularized by $R_{M,k}$ instead of $R_k$.

 Note that although it is not explicit from its definition, Eq.(\ref{eq_def_GammaA_LT}),  $\Gamma_{M,k}[\phi]$ is independent of $s$ as can be checked from Eq.(\ref{eq_def_GammaA_LT_nice}) or by deriving Eq.~\eqref{eq_def_GammaA_LT} with respect to $s$.

When evaluated in a constant field $\phi(\x)=s$ and at $k=0$,
\begin{equation}
\frac{\delta\Gamma_{M,k=0}}{\delta\phi(\x)}\bigg|_{\phi(\x)=s}=h
\end{equation}
 is a constant, by translation invariance, such that $\langle\hat\phi(\x)\rangle =s$. Thus, 
\begin{equation}
e^{-\Gamma_{M,k=0}[\phi(\x)=s]}=\int \DD\hat\phi\, e^{-{\cal H}(\hat\phi) -\frac {M^2}2[\int_\x(\hat\phi-s)]^2 +h\int_\x(\hat\phi-s)},
\end{equation}
and therefore
\begin{eqnarray}
\lim_{M\to\infty} e^{-\Gamma_{M,k=0}[\phi(\x)=s]}&\propto&\int \DD\hat\phi\, \delta\left(s-L^{-d}\int_\x \hat\phi\right)e^{-{\cal H}(\hat\phi)}\nonumber\\
&& \propto P(s).
\end{eqnarray}

This directly proves that $\lim_{M\to\infty}\Gamma_{M,k=0}[\phi(\x)=s]=L^d I(s)$. 
For finite $k$, it is convenient to define $\check \Gamma_k[\phi]=\lim_{M\to\infty}\Gamma_{M,k}[\phi]$, and we interpret $L^{-d}\check \Gamma_k[\phi(\x)=s]=I_k(s)$ as a scale-dependent rate function.

The similar construction at $M=0$ gives rise to the standard scale-dependent effective action $\Gamma_k[\phi]$ introduced in the FRG \cite{Dupuis2021}.

Note that at $k=0$, $\Gamma[\phi]=\Gamma_{k=0}[\phi]$ is  the (true) Legendre transform of $\ln \mathcal Z_{M=0}[h]$ (since the regulator $R_k$ identically vanishes in this limit) and is therefore a convex functional. On the contrary, $\check \Gamma[\phi]=\check\Gamma_{k=0}[\phi]$ is not a true Legendre transform even in this limit, due to the $M$ term in Eq.~\eqref{eq_def_GammaA_LT}, and can therefore be non-convex.

\section{Flow equations of the effective actions $\Gamma_{M,k}$ and Local Potential Approximation}

The initial condition of the flow corresponds to $k=\Lambda$, for which $\lim_{k\to\Lambda}R_k(\q)\to\infty$. A saddle-point analysis gives
\begin{equation}
\lim_{k\to\Lambda}\mathcal Z_{M,k}[h] \simeq e^{-\mathcal H_M[\phi] + h.\phi - \frac12 \phi.R_\Lambda.\phi},
\end{equation} 
 with $\phi[h]$ such that 
\begin{equation}
\frac{\delta\mathcal H_M}{\delta\phi(\x)} - h(\x) + \int_\y R_\Lambda(\x,\y)\phi(\y)=0.
\end{equation} 
Performing the Legendre transform and using Eq.~\eqref{eq_def_GammaA_LT}, we obtain
\begin{equation}
\lim_{k\to\infty}\Gamma_{M,k}[\phi]=\mathcal H[\phi].
\end{equation} 

The effective actions $\Gamma_{M,k}$ obey the exact flow equation
\begin{equation}
\partial_k \Gamma_{M,k}[\phi]=\frac{1}{2}\int_{\x,\y}\partial_k R_{M,k}(\x,\y) \left(\Gamma_{M,k}^{(2)}+R_{M,k}\right)^{-1}(\x,\y),
\end{equation} 
where 
\begin{equation}
\Gamma_{M,k}^{(2)}=\Gamma_{M,k}^{(2)}[\x,\y;\phi]=\frac{\delta^2\Gamma_{M,k} }{\delta\phi(\x)\delta\phi(\y)},
\end{equation} 
as can be derived by using the standard properties of Legendre transforms \cite{Dupuis2021}.

The Local Potential Approximation corresponds to the approximation
\begin{equation}
\Gamma_{M,k}[\phi]=\int_\x\left(\frac{(\partial\phi)^2}{2}+U_{M,k}(\phi(\x)\right),
\end{equation} 
from which follows the inverse propagator in momentum space and in constant field $\phi$
\begin{equation}
\Gamma_{M,k}^{(2)}(\q,\phi)=\q^2+U_{M,k}''(\phi).
\end{equation} 
In this approximation, the only flowing quantity is  $U_{M,k}$, and its flow equation reads
\begin{equation}
\partial_k U_{M,k}(\phi)=\frac{1}{2L^d}\sum_\q\frac{\partial_k R_{k}(\q)}{\q^2+R_k(\q)+M^2\delta_{\q,0}+U''_{M,k}(\phi)}.
\label{eq_LPA_dimful}
\end{equation} 
In the limit $M=0$, we recover the flow equation of the effective potential $U_k(m)$ (at finite size with periodic boundary conditions \cite{Fister2015})
\begin{equation}
\label{flow_u_lpa}
\partial_k U_{k}(m)=\frac{1}{2L^d}\sum_\q\frac{\partial_k R_{k}(\q)}{\q^2+R_k(\q)+U''_{k}(m)},
\end{equation} 
while in the limit $M\to\infty$ the zero-momentum contribution is suppressed, and the equation for $I_k$ reads
\begin{equation}
\label{flow_i_lpa}
\partial_k I_{k}(s)=\frac{1}{2L^d}\sum_{\q\neq 0}\frac{\partial_k R_{k}(\q)}{\q^2+R_k(\q)+I''_{k}(s)}.
\end{equation} 
In the thermodynamic limit $L\to\infty$, the two sums converge to the same integral, and we recover the equivalence between the two quantities, $I_k(s)=U_k(m=s)$.

To study the critical behavior, it is convenient to introduce the dimensionless fields $\tilde m=k^{-(d-2)/2}m$ and $\tilde s=k^{-(d-2)/2}s$, and  $\tilde U_{k}(\tilde\phi)=k^{-d}U_{k}(\tilde\phi k^{(d-2)/2})$ and $\tilde I_{k}(\tilde\phi)=k^{-d}I_{k}(\tilde\phi k^{(d-2)/2})$, as well as $\tilde \q=\q/k$. Writing the regulator as $R_k(\q)=k^2 \tilde R((\q/k)^2)$, the flows become
\begin{align}
k\partial_k \tilde U_{k}=&-d \tilde U_{k}+\frac{d-2}{2}\tilde m\tilde U_{k}'+ \frac{1}{2 (kL)^d}\sum_{\tilde\q}\frac{2\tilde R-\tilde\q^2 \tilde R' }{\tilde\q^2+\tilde R+\tilde U''_{k}},
\label{eq_LPA_dimless}
\end{align} 
and
\begin{align}
k\partial_k \tilde I_{k}=&-d \tilde I_{k}+\frac{d-2}{2}\tilde s\tilde I_{k}'
+ \frac{1}{2 (kL)^d}\sum_{\tilde\q\neq 0}\frac{2\tilde R-\tilde\q^2 \tilde R' }{\tilde\q^2+\tilde R+\tilde I''_{k}}.
\label{eq_LPA_dimless2}
\end{align} 

In the thermodynamic limit, both flow equations become the standard dimensionless LPA equation
\begin{equation}
k\partial_k \tilde U_{k}=-d \tilde U_{k}+\frac{d-2}{2}\tilde m\tilde U_{k}'+ \frac{1}{2}\int_{\tilde \q}\frac{2\tilde R-\tilde\q^2 \tilde R' }{\tilde\q^2+\tilde R+\tilde U''_{k}},
\label{eq_LPA_thermo}
\end{equation} 
and
\begin{equation}
k\partial_k \tilde I_{k}=-d \tilde I_{k}+\frac{d-2}{2}\tilde sI_{k}'+ \frac{1}{2}\int_{\tilde \q}\frac{2\tilde R-\tilde\q^2 \tilde R' }{\tilde\q^2+\tilde R+\tilde I_{k}''},
\label{eq_LPA_thermo2}
\end{equation} 
and at criticality, the dimensionless effective potential reaches a fixed point as $k\to 0$, that is, $\lim_{k\to 0}\lim_{L\to \infty}\tilde U_{k}=\lim_{k\to 0}\lim_{L\to \infty}\tilde I_{k}=\tilde U^*$, with $\tilde U^*$ the solution to
\begin{equation}
0=-d \tilde U^*+\frac{d-2}{2}\tilde\phi\tilde U^{*'}+ \frac{1}{2}\int_{\tilde \q}\frac{2\tilde R-\tilde\q^2 \tilde R' }{\tilde\q^2+\tilde R+\tilde U^{*''}}.
\label{eq-FP}
\end{equation} 
At large field, the last term can be neglected in Eq. \eqref{eq-FP}, and $\tilde U^*$ behaves as a power law:
\begin{equation}
\tilde U^*\propto \tilde\phi^{\frac{2d}{d-2+\eta}},
\end{equation} 
with $\eta=0$ at LPA.

At finite size $L$ and at $T=T_c$, the flow equations \eqref{eq_LPA_dimless} and \eqref{eq_LPA_dimless2} are indistinguishable from Eqs.~\eqref{eq_LPA_thermo} and \eqref{eq_LPA_thermo2} as long as $k L\gg 1$, and $\tilde U_k$ and $\tilde I_{k}$ go to the same fixed point solution $\tilde U^*$: For $a^{-1}\gg k\gg L^{-1}$,  $I_k(s)\simeq U_k(m=s)\simeq k^{d}\tilde U^*(k^{-(d-2)/2}s)$, which corresponds to the correct scaling $(d-2+\eta)/2$ with $\eta=0$ at LPA.

However, for $kL\lesssim 1$, the flows of the $U_{M,k}$ differ significantly. In particular, the flow of the effective potential is 
\begin{equation}
\partial_k U_{k}(m)\simeq \frac{1}{2L^d}\frac{\partial_k R_{k}(0)}{R_k(0)+U''_{k}(m)},
\end{equation} 
which corresponds to the flow of a 0-dimensional theory. Because $R_k(0)\to0$ as $k\to 0$, $U''_{k}(m)$ cannot stay negative in the denominator: This induces a return to convexity, as expected for a (true) Legendre transform.

On the other hand, the flow of the rate function stops very quickly for $k\lesssim 1/L$ since the contribution of the finite momenta is negligible thanks to $\partial_k R_{k}(\q)$ in this regime.

Finally, at large field $\phi$, the flow of $U_{M,k}$ is barely modified by the finite size effects, and we recover the powerlaw behavior of the effective potential and rate function, $U_{M,k=0}\propto \phi^{\frac{2d}{d-2+\eta}}$ (with $\eta=0$ at LPA).

\section{Monte Carlo simulations}

The PDFs were estimated numerically using Monte-Carlo simulations of the Ising model on the cubic lattice with periodic boundary conditions, using Wolff's algorithm \cite{Wolff1989}. 
The correlation length can be estimated using \cite{Lundow2018}
\begin{equation}
    \xi_{\infty,{\rm MC}}/a = \frac{1.073}{\sqrt{T/J}}\left(1-\frac{T_c}{T}\right)^{-\nu},
\end{equation}
with $J/T_c\simeq 0.221654626$ \cite{Ferrenberg2018} and $\nu=0.62997$ \cite{Kos2016}. To compute the PDF for various $L$ and fixed $\zeta$,  the temperature is therefore chosen such that $\xi_{\infty,{\rm MC}}=L/\zeta$. 
We performed simulations for linear sizes from $L=16$ to $L=128$, with for each size at least $1.5\times 10^7$ measurements of the total spin and energy. The latter allows for using histogram reweighting \cite{Ferrenberg1988} to explore several values of $\zeta$ in one simulation.
In practice, we performed simulations for each $L$ at $\zeta=-3,0,3$.

\begin{figure}[!t]
    \centering
    \includegraphics[width=8cm]{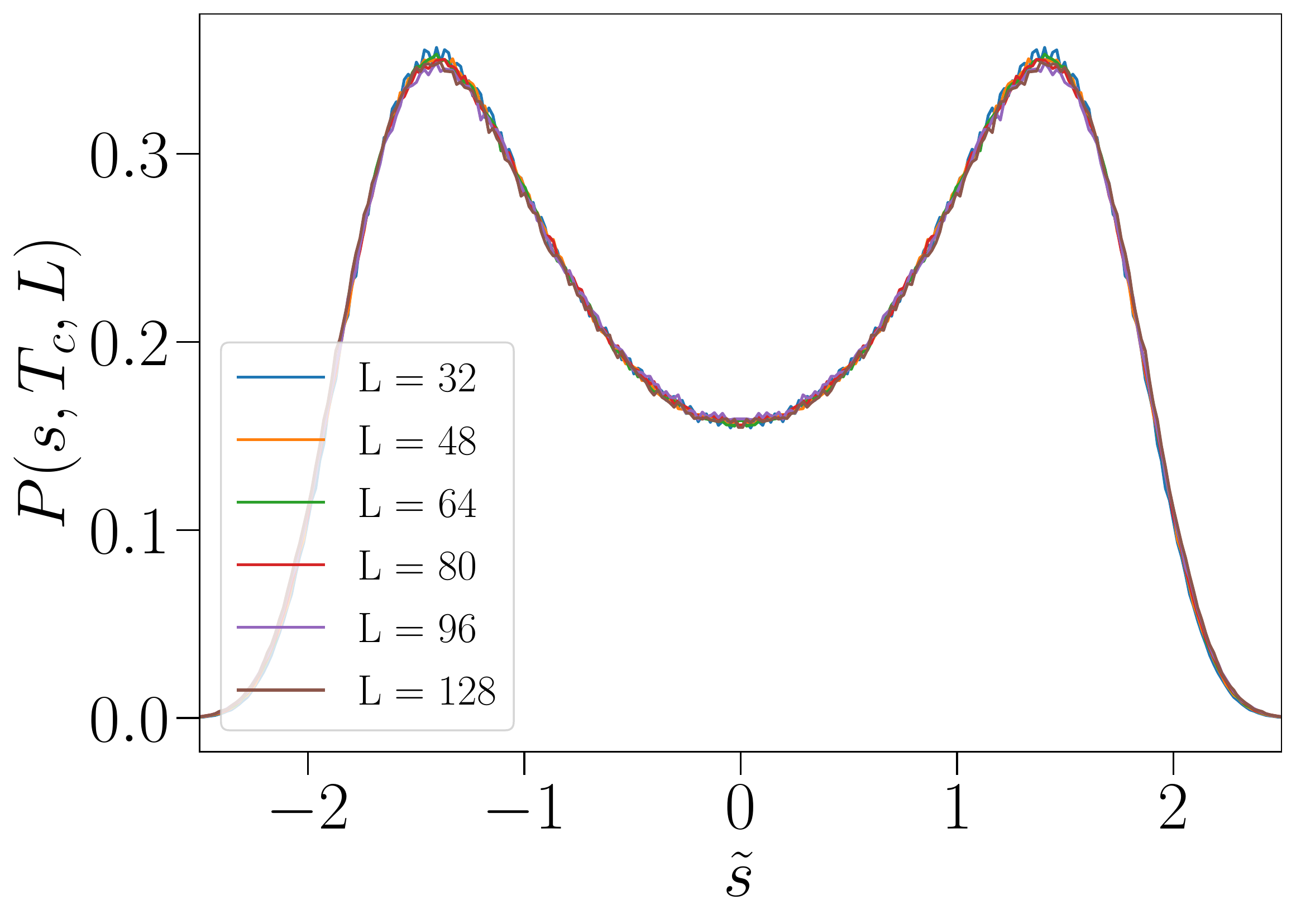}
    \caption{Probability distribution at $T_c$ obtained from Monte-Carlo simulations for various sizes. The finite size effects are very weak.}
    \label{fig_finite_size_MC}
\end{figure}

The rate function is given by
\begin{equation}
    I(s,\xi_{\infty},L) = \frac{-1}{L^3}\ln P(s,\xi_{\infty},L).
\end{equation}
At criticality, the scaling form is obtained from
\begin{equation}
    I_\zeta(\tilde s) = L^3 I(\tilde s L^{-\frac{d-2+\eta}{2}},\xi_{\infty},L)=-\ln P(\tilde s L^{-\frac{d-2+\eta}{2}},\xi_{\infty},L).
\end{equation}
Therefore, when expressed in terms of $\tilde s$, the probability distribution is expected to converge to a limit form, at least for sizes large enough. This happens already for moderate sizes, as finite size effects are very weak, as can be seen from Fig.~\ref{fig_finite_size_MC}. To obtain such a figure, for each length, the measured total spin is rescaled  by $L^{\frac{d-2+\eta}{2}}$, with $\eta=0.036298$ \cite{Kos2016}, and  the data are binned into approximately 150 bins. Since the finite size effects are small, when comparing to FRG, we use the data at $L=128$. 

It is important to realize that the universality of the scaling functions $I_\zeta(\tilde s)$ does not mean that all systems of the same universality class share the same rate functions, but only  that they do so once the length and field normalizations have been fixed. All rate functions of a given universality class are therefore of the form $\alpha I_\zeta(\beta\tilde s)$, with $\alpha$ and $\beta$ system-dependent numbers. Therefore, the comparison between two rate functions obtained for instance in the Ising model and  from FRG, is made up to two normalizations that can be chosen for instance such that they coincide at one point. We stress that $\alpha$ and $\beta$ do not depend on $\zeta$, and only have to be determined once.

To determine $\beta$, associated with the magnitude of the microscopic degrees of freedom, we impose that the rate function at $\zeta=0$ has its (right) minimum at $1$. For this, we locate the position of the minimum, noted $\tilde s_0$. It is   very weakly dependent on $L$, and we therefore use its value at $L=128$.

The amplitude $\alpha$ is associated with a length scale. Note that we have chosen to shift the rate functions such that they vanish at $s=0$. This is always feasible by changing the pre-factor of the PDF which is ultimately determined by imposing that the PDF is normalized. Therefore, it is possible to determine the amplitude $\alpha$ by imposing that $I_{\zeta=0}(1) =-1$. Once the two amplitudes have been determined, they are used to rescale the total spin and rate functions for all values $\zeta$.

\begin{figure}[!t]
    \centering
    \includegraphics[width=8cm]{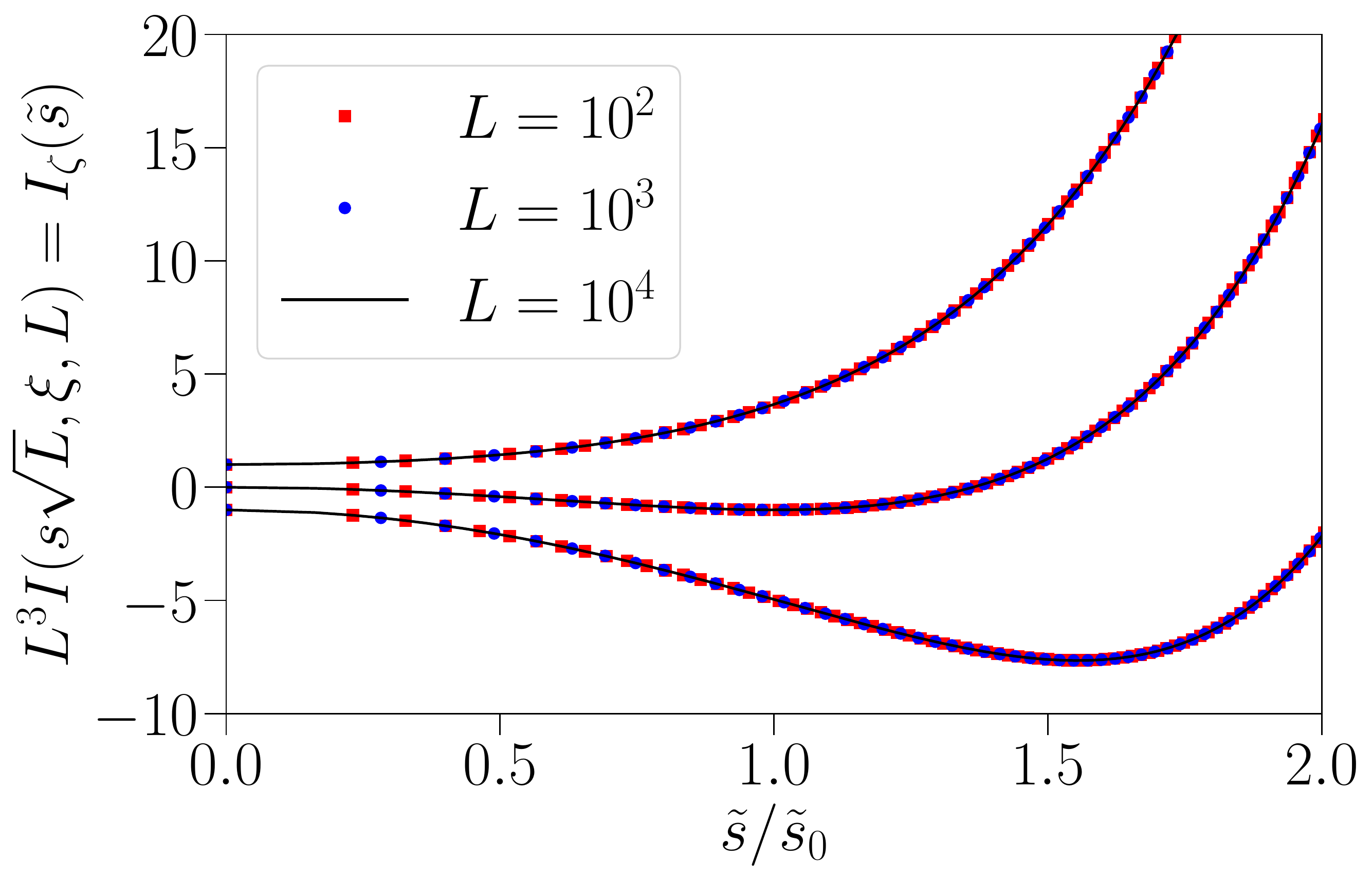}
    \caption{Rate functions rescaled by $L^d$ as a function of $\tilde s$ at $\zeta=3$ (top), $\zeta=0$ (middle) and $\zeta=-3$ (bottom) obtained from FRG with the ``exponential regulator'' $R_k(\q)=\alpha k^2 e^{-\q^2/k^2}$ and $\alpha=4.65$. For each value of $\zeta$, the solid line shows the results for $L=10^4$ in units of a mesoscopic scale $k_*^{-1}$ at which the system becomes scale invariant. Symbols are for $L=10^3$ (circles) and $10^2$ (squares). The rate functions have been normalized as explained in the main text. This shows that the rate function obeys the expected scaling form $I(s,\xi_{\infty},L)=L^{-3}I_{\zeta}(\tilde s)$. }
    \label{fig_scalingproof}
\end{figure}

\section{Numerical integration of the LPA flow equation}

\subsection{Numerical implementation}

For the numerical integration of partial differential equations, Eqs.~\eqref{flow_u_lpa} and \eqref{flow_i_lpa}, we use the Euler method with RG time steps of $10^{-4}$ with $l=\ln(k/k_*)$ the RG time (integration with Runge-Kutta of order 4 gives similar results). Here $k_*^{-1}$ is a mesoscopic length scale for which the system is self-similar, i.e. a scale much larger than the lattice spacing and at which the effective potential and rate function are described by the fixed point potential. Length scales are measured in units of  $k_*^{-1}$.

The field dependence of functions $U$ and $I$ is discretized on a grid of $200$ points, with the position of the minimum of the initial condition at roughly $1/4$-$1/5$ of the grid.
To integrate the flow Eqs.~\eqref{flow_u_lpa} and \eqref{flow_i_lpa} in practice, we run the flow in terms of dimensionless quantities, i.e. integrate Eq.~\eqref{eq_LPA_dimless}, until $\min(kL,k\xi_{\infty})\simeq 4$. Then we switch to dimensionful quantities, i.e. integrate Eq.~\eqref{eq_LPA_dimful}, and run the flow until termination. 
For finite $L$, the sum over discrete momenta can be very demanding numerically. Therefore, we do numerical integration over the continuous variable $q/k$ instead of sums over discrete momenta  until $kL\simeq 40$. By varying all parameters of our numerical integration of the flows, we have checked that our results are converged.

\subsection{Determination of $\zeta$}
To compare the FRG results to the MC simulations, it is necessary to compute $\zeta$ for various RG flow initial conditions. At LPA, the correlation length in the disordered phase and in the thermodynamic limit is obtained from \cite{Dupuis2021}
\begin{equation}
    \xi_{\infty,{\rm FRG}}^2=\lim_{L\to\infty}\frac{1}{U''_{k=0}(m=0)}. 
\end{equation}
In practice, we solve the flow of the  effective potential of an infinite system, Eq.~\eqref{eq_LPA_thermo} with $M=0$, starting from an initial condition which is the fixed point potential $\tilde U^*$ perturbed by a quadratic term $\delta,\tilde\phi^2/2$. For $\delta=0$, the system is critical, while it flows to the disordered (ordered) phase for $\delta>0$ ($\delta<0$). For small $\delta$, the correlation length behaves as
\begin{equation}
    \xi_{\infty,{\rm FRG}}\approx \xi_+ \delta^{-\nu_{\rm LPA}},
\end{equation}
with $\nu_{\rm LPA}\simeq 0.65$, (instead of the ``exact'' $\nu=0.62997$ from conformal bootstrap \cite{Kos2016}), and $\xi_+$ is a non-universal number. Note that $\nu_{\rm LPA}$ and $\xi_+$ depend slightly on the regulator, which we take into the account to determine $\xi_{\infty,FRG}$.

\begin{figure}[!t]
    \centering
    \includegraphics[width=8cm]{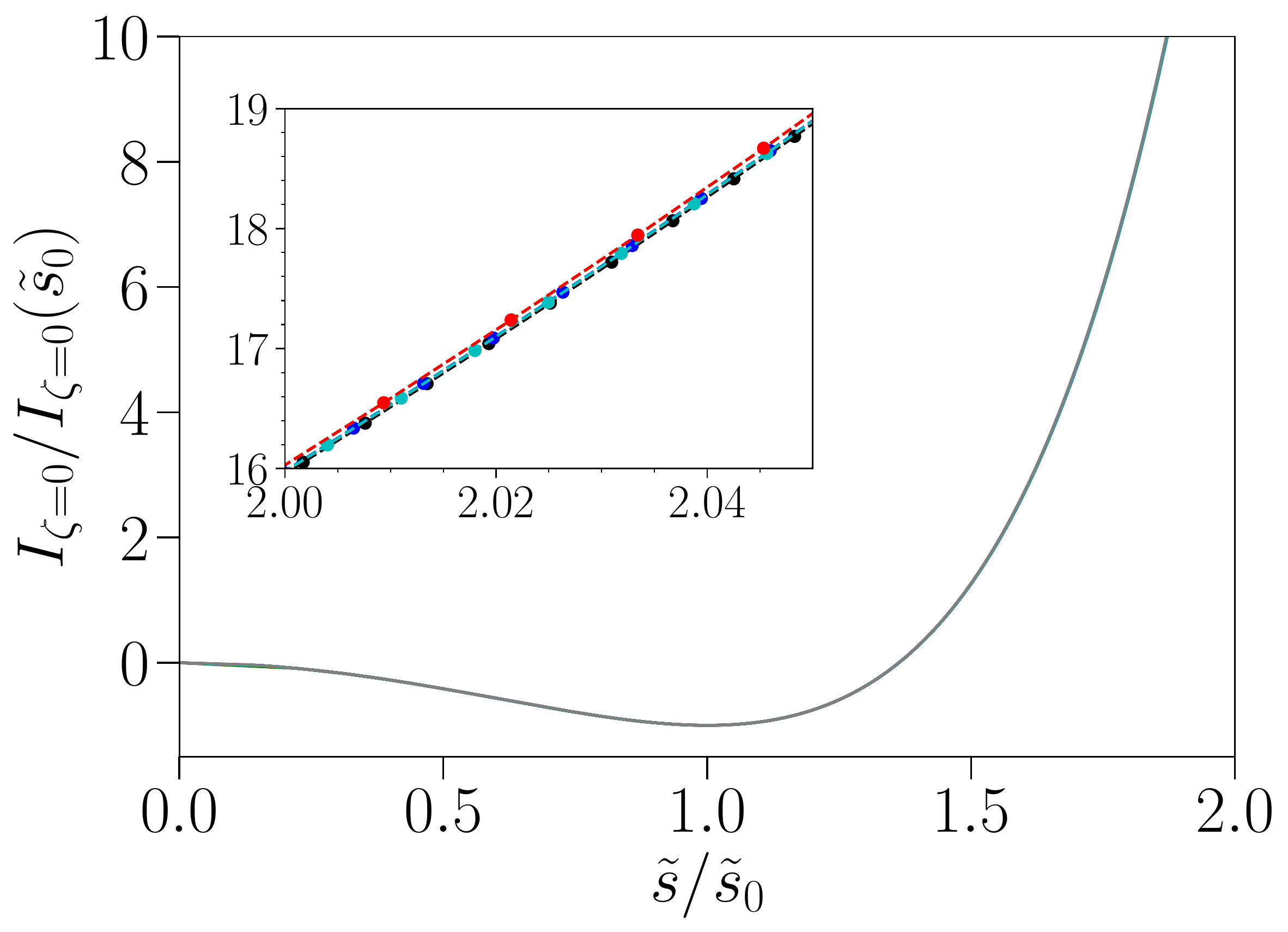}
    \caption{Rate functions at $\zeta=0$  obtained from FRG with the ``exponential regulator'' $R_k(\q)=\alpha k^2 e^{-\q^2/k^2}$ with $\alpha=0.75$ (magenta), $\alpha=2.5$ (black), $\alpha=4.65$ (blue), the ``Litim regulator'' $R_k(\q)=\alpha (k^2-\q^2) \theta(k^2-\q^2)$ with $\alpha=1$ (red) and $\alpha=2$ (green), and the ``Wetterich regulator'' $R_k(\q)=\alpha \q^2/(\exp(\q^2/k^2)-1)$ with $\alpha=0.75$ (gray), $\alpha=6.05$ (cyan). The rate functions have been normalized as explained in the main text. The different results are almost indistinguishable. The inset is a zoom in the tail, showing the same data (with symbols and dashed line) and the very weak spread induced by the regulator dependence. }
    \label{fig_Rdependence}
\end{figure}

To compute the rate function for a given $\zeta_{FRG}\geq0$, at a given $L$, the procedure is the following. We first solve the flow of the effective potential in the thermodynamic limit, tuning $\delta$ such that the correlation length obtained from the flow is equal to $L/\zeta_{FRG}$. Then, we compute the rate function by solving the flow (with finite $L$) for the same initial condition (recall that for $kL\gg1$, the flows of $I_k$ and $U_k$ are identical, and that they have the same initial condition). We have checked that our results obey scaling, i.e. changing the couple $(L,\delta)$ at fixed $\zeta_{{\rm LPA}}$ gives the same solution, see Fig.~\ref{fig_scalingproof}. In our comparison with MC simulations, we choose $L=10000$.
For negative $\zeta_{FRG}$, the initial condition is the value of $\delta$ (which is negative) such that $\xi_{\infty,FRG}(-\delta)=L/|\zeta_{FRG}|$.

\begin{figure}[!t]
    \centering
    \includegraphics[width=8cm]{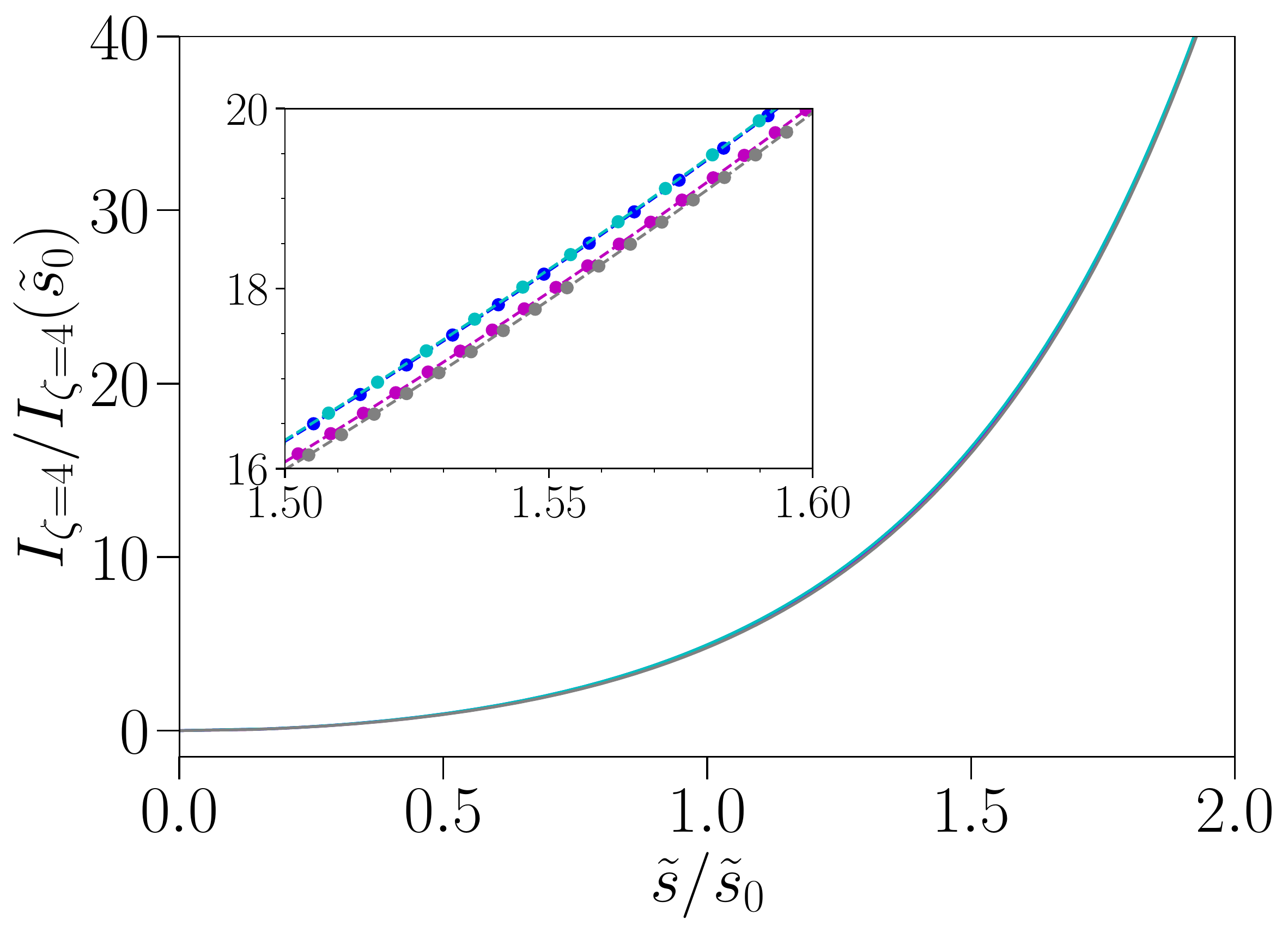}
    \caption{Rate functions at $\zeta=4$  obtained from FRG with the ``exponential regulator'' $R_k(\q)=\alpha k^2 e^{-\q^2/k^2}$ with $\alpha=0.75$ (magenta), $\alpha=4.65$ (blue), and the ``Wetterich regulator'' $R_k(\q)=\alpha \q^2/(\exp(\q^2/k^2)-1)$ with $\alpha=0.75$ (gray), $\alpha=6.05$ (cyan). The rate functions have been normalized using the normalization obtained for $\zeta=0$ with the same regulator (see Fig.~\ref{fig_Rdependence}).  The inset is a zoom in the tail, showing the same data (with symbols and dashed line) and the weak spread induced by the regulator dependence, see inset. }
    \label{fig_Rdependence_zeta4}
\end{figure}

\subsection{Regulator dependence}

The results showed in the main text were obtained by solving the LPA equations using the ``exponential regulator'' $R_k(\q)=\alpha k^2 e^{-\q^2/k^2}$, where $\alpha =4.65$ is the optimized value at LPA for this regulator (i.e. the value obtained from the Principle of Minimal Sensitivity, for which the critical exponent $\nu$ varies the least when changing $\alpha$, see \cite{Balog2019} for details). The dependence of the rate function on the regulator is very weak, as observed in Fig.~\ref{fig_Rdependence} for $\zeta=0$ and Fig.~\ref{fig_Rdependence_zeta4} for $\zeta=4$. 

\begin{figure}[t]
    \centering
    \includegraphics[width=8cm]{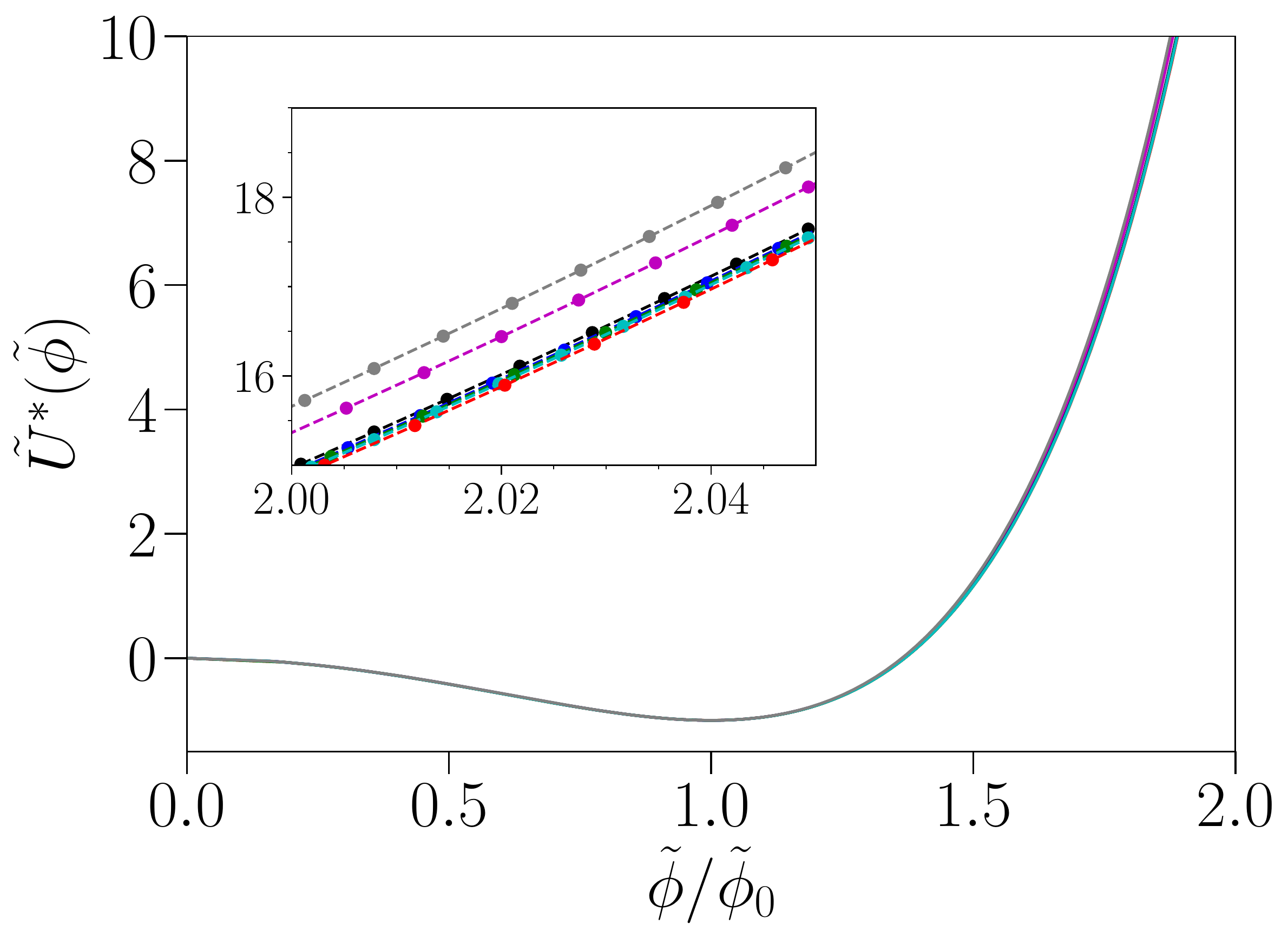}
    \caption{Fixed point potentials obtained from FRG with the ``exponential regulator'' $R_k(\q)=\alpha k^2 e^{-\q^2/k^2}$ with $\alpha=0.75$ (magenta), $\alpha=2.5$ (black), $\alpha=4.65$ (blue), the ``Litim regulator'' $R_k(\q)=\alpha (k^2-\q^2) \theta(k^2-\q^2)$ with $\alpha=1$ (red) and $\alpha=2$ (green), and the ``Wetterich regulator'' $R_k(\q)=\alpha \q^2/(\exp(\q^2/k^2)-1)$ with $\alpha=0.75$ (gray), $\alpha=6.05$ (cyan). They have been normalized such that $\tilde U^*(\tilde\phi_0)=-1$, with $\tilde\phi_0$ the position of the minimum. There is only a weak dependence on the regulator after rescaling, only showing in the tail, see inset. }
    \label{fig_Rdependence_FP}
\end{figure}

In Fig.~\ref{fig_Rdependence_FP}, we show the regulator dependence of the fixed point effective potential. While it is expected to depend on the regulator, quite surprisingly, we observe that after normalization, the dependence is very weak. Note that this weak dependence is not an artifact of the LPA, since it can be shown in the large $N$ limit of the $O(N)$ model, where the LPA is exact, that this dependence  is still present\cite{next_paper_2035}.

Let us remark that the regulator functions we use in practice (see caption of Fig.~\ref{fig_Rdependence_FP}) do not formally diverge at the beginning of the flow. 
This implies in practice that the initial condition of the flow is not strictly given by the mean-field approximation (i.e. $\Gamma_{M,k\sim a^{-1}}$ is  not exactly given by $\mathcal H$). This is however irrelevant for the problem studied here, since we are only interested in the universal features of the PDF, which by definition are independent of the microscopic details of the model: running the flow of $\Gamma_{M,k}$ with initial conditions either given by ${\cal H}$ or obtained with a finite regulator function yields the same fixed point and scaling function $I_{\zeta}(\tilde s)$.

\subsection{Comparison with Monte-Carlo simulations}
To compare our FRG calculations to the simulations, we normalize the rate functions as explained above, i.e. we impose that at $\zeta=0$ the rate function is $-1$ at $1$. We use the same normalization (obtained for $\zeta=0$) to normalize the rate function for the other values of $\zeta$.

We have observed that if we use $\zeta_{FRG}=\zeta_{ {\rm MC}}$, we do not obtain a collapse of our FRG calculations with the simulations. However, allowing for a global rescaling $\zeta_{FRG}\simeq\zeta_{ MC}/0.9$, we find a very good agreement between the two methods. We attribute that to the fact that the LPA is not exact, and that therefore the correlation length obtained from FRG is slightly off.


\end{document}